\documentclass[aps,prd, amsfonts, amssymb, amsmath, superscriptaddress, notitlepage,twocolumn,10pt,nobalancelastpage]{revtex4-1}
\usepackage{graphicx} 
\usepackage{hyperref}
\usepackage{bm}
\hypersetup{ 
    colorlinks=true, 
    linktoc=all,     
    linkcolor=blue,  
    citecolor =blue
} 
\usepackage{amsmath}

\renewcommand{\v}[1]{\bm{#1}} 
\newcommand{\avg}[1]{\langle#1\rangle} 
\newcommand{\abs}[1]{\left| #1 \right|} 

\begin{document}

\title{Fermionic criticality of anisotropic nodal point semimetals away from the upper critical dimension: exact $1/N_f$ exponents}

\author{Mikolaj D. Uryszek}
\affiliation{London Centre for Nanotechnology, University College London, Gordon St., London, WC1H 0AH, United Kingdom}
\author{Frank Kr\"uger}
\affiliation{London Centre for Nanotechnology, University College London, Gordon St., London, WC1H 0AH, United Kingdom}
\affiliation{ISIS Facility, Rutherford Appleton Laboratory, Chilton, Didcot, Oxfordshire OX11 0QX, United Kingdom}
\author{Elliot Christou}
\affiliation{London Centre for Nanotechnology, University College London, Gordon St., London, WC1H 0AH, United Kingdom}

\begin{abstract}
We consider the fermionic quantum criticality of anisotropic nodal point semimetals in $d = d_L + d_Q$ spatial dimensions that disperse linearly in $d_L$ dimensions, 
and quadratically in the remaining $d_Q$ dimensions. When subject to strong interactions, these systems are susceptible to semimetal-insulator transitions concurrent 
with spontaneous symmetry breaking. Such quantum critical points are described by effective field theories of anisotropic nodal fermions coupled to dynamical order 
parameter fields. We analyze the universal scaling in the physically relevant spatial dimensions, generalizing to a large number $N_f$ of fermion flavors for analytic 
control. Landau damping by gapless fermionic excitations gives rise to non-analytic self-energy corrections to the bosonic order-parameter propagator that dominate 
the long-wavelength behavior. We show that perturbative momentum shell RG leads to non-universal, cutoff dependent results, as it does not correctly account for this non-analytic structure.
In turn, using a completely general soft cutoff formulation, we demonstrate that the correct IR scaling of the dressed bosonic propagator can be deduced by
enforcing that results are independent of the cutoff scheme. Using this soft cutoff approach, we compute the exact critical exponents  for anisotropic semi-Dirac fermions 
$(d_L=1$, $d_Q=1)$ to leading order in $1/N_f$,  and to all loop orders. Applying the same method to relativistic Dirac fermions, we reproduce the critical 
exponents  obtained by other methods, such as conformal bootstrap. Unlike in the relativistic case, where the UV-IR connection is re-established at the upper critical dimension, 
non-analytic IR contributions persist near the upper critical line $2 d_L+d_Q=4$ of anisotropic nodal fermions. 
We present $\epsilon$-expansions in both the number of linear and quadratic dimensions. The corrections to critical exponents are non-analytic in $\epsilon$, with a functional 
form that depends on the starting point on the upper critical line. 
\end{abstract}

\maketitle

\section{Introduction}

The discovery of topological insulators has led to an explosion of research into topological aspects of electronic band structures in two and three dimensions \cite{hasan_kane_top_insulators_review,zhang_top_ins_review}. In so-called nodal point semimetals, valence and conduction bands touch at a number of 
discrete points in the Brillouin zone. The most fundamental members of this family are Weyl or Dirac semimetals \cite{armitage_review_semimetals,yan_felser_weyl_SM_review,vafek_vishwanath_diracF_in_materials_review}, 
which exhibit relativistic low-energy excitations that are protected by topology and symmetry. 

\begin{figure}[t]
 \includegraphics[width=1.0\columnwidth]{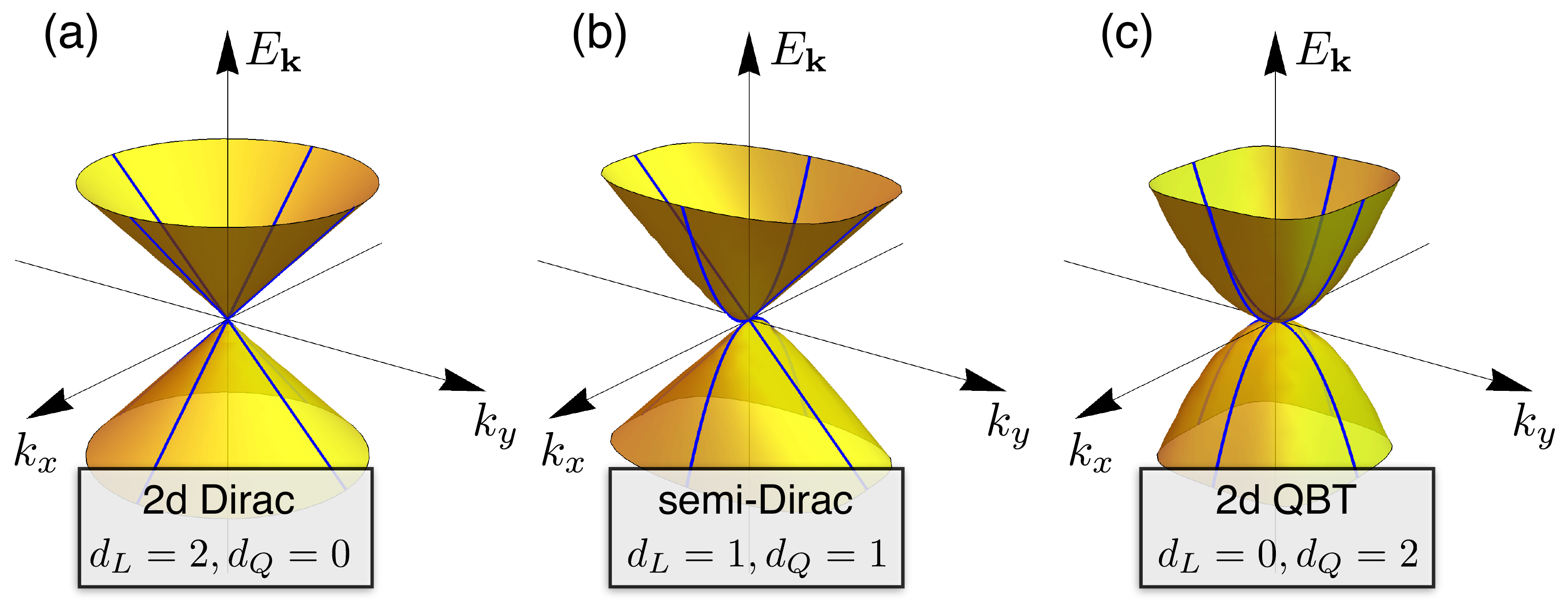}
 \caption{Different types of nodal-point semimetals in $d=2$ spatial dimensions. The quasiparticles at the band-touching point disperse linearly along $d_L$ and quadratically along $d_Q$ 
 directions, $d_L+d_Q=2$.}
\label{fig1}
\end{figure}

A transition into a gapped insulating state can only be achieved by breaking the protecting symmetry or by tuning the band structure through a topological phase transition where 
nodal points with opposite chirality merge. Such a topological phase transition was observed in black phosphorous \cite{kim_SD_black_Phos,rodin_SD_black_phos}, and 
is predicted to occur in strained honeycomb lattices \cite{pardo_pickett_SD_strained_honeycomb} and VO$_2$--TiO$_2$ heterostructures \cite{pardo_pickett_SD_strained_honeycomb,SD_tivo2}. 
At the transition point, the dispersion becomes quadratic along the momentum direction along which the nodal points merge, whilst it remains relativistic along the other direction \cite{monathmaux_semi_dirac,montambaux_sd_merging}. Such quasiparticles were termed semi-Dirac fermions \cite{pardo_pickett_SD_strained_honeycomb,pardo_SD}.
Analogous hybrid quasiparticles exist at topological quantum phase transitions in non-centrosymmetric three-dimensional materials \cite{nagaosa_top_QPTs_3D,nagaosa_semi_dirac_nature}.  
Anisotropic nodal fermions  with $d_L$ linear and $d_Q$ quadratic momentum directions  in $d=d_L+d_Q$ spatial dimensions interpolate between relativistic Dirac or Weyl fermions  and quasiparticles 
in systems with quadratic band touching \cite{bilayer_graphene_review,fu_top_crys_insulator_QBT,balents_pyrchlore_iridates_QBT,balents_soc_review_QBT} (see Fig.~\ref{fig1}).
 
In anisotropic nodal point semimetals the density of states vanishes with a modified power law, $\rho(E)\sim |E|^r$, $r=(2 d_L+d_Q-2)/2$, giving rise to new exponents in the temperature dependence of various thermodynamic quantities, such as specific heat and compressibility \cite{nagaosa_top_QPTs_3D}. 
Moreover, the strong anisotropy due to linear and quadratic momentum directions leads to unusual anisotropic transport phenomena \cite{nagaosa_top_QPTs_3D,Link+18} and exotic, directionally dependent screening effects \cite{nagaosa_semi_dirac_nature,nagaosa_semi_dirac_prl,Cho+16,Han+19}.

Nodal semimetals with point-like Fermi surfaces provide the simplest setting to study fermionic quantum criticality. 
Quantum phase transitions can be driven by sufficiently strong short-range electron-electron interactions in the underlying lattice model. 
Depending on the nature of the microscopic interactions, this can lead to various types of symmetry breaking, resulting in rich phase diagrams with antiferromagnetic, charge-density wave, and bond-ordered phases, as studied in great detail for extended Hubbard models on the honeycomb lattice \cite{Grushin+13,Garcia+13,Daghofer+14,Capponi+15,Motruk+15,Scherer+15,Volpez+16,Kurita+16,Pena+17,Christou+18}. 
Irrespective of the particular order, the spontaneous symmetry breaking generically leads to the opening of a gap in the fermion spectrum and is therefore concurrent with  semimetal-insulator transitions.

The universality of a particular transition can be studied using an effective field theory that is derived through a Hubbard-Stratonovich decoupling of the interaction vertex in the 
relevant channel, followed by the conventional coarse graining procedure. This results in a dynamical bosonic order-parameter theory which is coupled to the gapless fermion excitations. 
In the purely relativistic case of Dirac fermions, this is known as the Gross-Neveu-Yukawa (GNY) theory \cite{herbut_graphene_2006,herbut_roy_honeycomb,Assaad+13}, which describes chiral 
symmetry breaking and spontaneous mass generation in high-energy physics \cite{Gross+74,ZinnJustin91}.  The coupling between the order parameter fields and the gapless Dirac fermions 
leads to novel fermion-induced critical behavior that falls outside the Landau-Ginzburg-Wilson paradigm of a pure order parameter description \cite{Li+17}. 

The lack of Lorentz invariance and the different scaling of the density of states near the nodal points leads to distinct fermion-induced criticality in nodal fermion systems 
with quadratic \cite{lemonik_sym_breaking_bilayer_graphene,herbut_janssen_qbt_top_mott_QBT,janssen_herbut_qbt_nematic,vafek_bilayer_rg} and semi-Dirac \cite{Wang+17,Roy+18,uryszek_semi_dirac,sur_roy_semi_dirac} band-touching points. The latter are particularly interesting because the intrinsic electronic anisotropy gives rise to highly anisotropic 
order-parameter correlations with different correlation-length exponents along linear and quadratic momentum directions. 

At the same time, the  anisotropic dispersion of semi-Dirac fermions makes this problem  difficult. Different complementary expansions were used to obtain analytic control 
in renormalization-group (RG) calculations. In Ref.~\cite{Roy+18}, the problem was analyzed in two spatial dimensions but with a generalized dispersion $k_x^{2n}$ in the non-relativistic 
direction, facilitating a controlled ascent from one dimension ($n \to\infty$). More traditional approaches include a $1/N_f$ expansion in the number of fermion flavors \cite{uryszek_semi_dirac}
and an $\epsilon$ expansion below the line of upper critical dimensions $2d_L +d_Q = 4$, expanding in the number of quadratic directions, $d_L=1$, $d_Q=2-\epsilon_Q$ \cite{sur_roy_semi_dirac}.

Under conventional momentum-shell RG, the bosonic order-parameter propagator develops unphysical divergencies, irrespective of the expansion 
scheme \cite{sur_roy_semi_dirac,uryszek_semi_dirac}.  This is because along the linear
momentum directions, the loop corrections to the propagator, obtained by successive integration of modes from a shell near the UV cutoff,  are irrelevant in an RG sense. The related 
divergencies need to be regularized by an additional IR contribution to the bosonic propagator that is not generated or renormalized under the Wilsonian RG. Instead it needs to be computed 
separately by integrating the fermion polarization diagram over the entire frequency and momentum range up to the infinitesimal 
shell \cite{nagaosa_semi_dirac_nature,Han+19,sur_roy_semi_dirac,uryszek_semi_dirac,Li+18,nagaosa_semi_dirac_prl,Cho+16}. 

As explained in Ref.~\cite{nagaosa_semi_dirac_prl}, the correct procedure within the large $N_f$ formulation is to compute bosonic and fermionic self-energies in a 
self-consistent scheme and to use
the dressed dynamical propagators as input in subsequent RG calculations. Such an approach was commonly used to understand quantum critical behavior 
of metals \cite{Altshuler+94,Polchinski94,Abanov+03,dtSon_coulomb_graphene,Metlitski+10,mross_senthil_rpa_fermi_surface,Efetov+13,Chubukov+14},
although it  
was later shown that the $1/N_f$ expansion fails at higher-loop order in systems with a full two-dimensional Fermi surface \cite{ssLee_fermi_surf_2+1}.

The necessity to use a dressed boson propagator is not a mere technical issue.  It is intimately linked to the phenomenon of Landau damping of 
order-parameter fluctuations by gapless electronic particle-hole excitations. This damping is known to completely change the long-wavelength behavior of the system, leading to distinct 
critical behavior. In itinerant ferromagnets, long-range spatial correlations associated with the Landau damping of the order parameter field generate a negative, non-analytic 
contribution to the static magnetic susceptibility, rendering the Hertz-Millis-Moriya theory \cite{hertz_quantum_critical_phenomena,milllis_nonzero_ferm_qcp} unstable towards 
first-order behavior or incommensurate order \cite{Chubukov+04}.

In order to identify the universal critical behavior of a general $d_L$-$d_Q$ nodal-fermion system, it is of crucial importance to use the correct bosonic IR propagator. 
However, due to the inherent anisotropy, the evaluation of the fermionic polarization diagram that determines the bosonic self energy $\Pi(\v{q},\Omega)$ is rather 
involved \cite{nagaosa_semi_dirac_prl}. As one might anticipate, 
$\Pi(\v{q},\Omega)$ is non-analytic and highly anisotropic, and often approximations or interpolations between different asymptotic forms are
used \cite{nagaosa_semi_dirac_prl,Cho+16,uryszek_semi_dirac,sur_roy_semi_dirac}, potentially leading to non-universal results. This problem is apparent in
recent studies of the effects of long-range Coulomb interactions between semi-Dirac fermions \cite{nagaosa_semi_dirac_prl,Cho+16}. 
While the Coulomb interaction in two dimensions is represented by a bare gauge-boson propagator $G_\phi^{-1}\sim |\v{q}|$, the 
long-wavelength behavior is completely dominated by the non-analytic bosonic self energy $\Pi(\v{q},\Omega)$, giving rise to marginal Fermi-liquid behavior at smallest energies,  
with various anomalous physical properties \cite{nagaosa_semi_dirac_prl}.  Using an incomplete IR propagator, e.g. neglecting the dynamic part of $\Pi(\v{q},\Omega)$, leads to 
fundamentally different results \cite{Cho+16}.

In this article we revisit the quantum criticality of topological nodal-point semimetals, using a soft cutoff RG approach 
\cite{brezin_zinn_justin_book,vojta_sachdev_dwave_field_theoretic_derivative,vojta_zhang_sachdev_d_wave_SC,Vojta+08,huh_sachdev_nematic_d_wave_SC} within the large $N_f$ expansion. 
We demonstrate that the use of an incorrect bosonic IR propagator
gives rise to non-universal results that depend on the UV cutoff scheme. In turn, enforcement of cutoff independence leads to the correct scaling form of the bosonic
IR propagator, which is given by the full RPA fermion loop re-summation. 

Using the soft cutoff approach with the dressed order-parameter propagator, we compute the exact critical 
exponents for anisotropic semi-Dirac ($d_L=1$, $d_Q=1$) 
and relativistic Dirac ($1<d=d_L<3$) fermions to leading order in $1/N_f$,  and to all loop orders. We include the isotropic relativistic case to illustrate the problem of
unphysical cutoff dependence, to introduce the methodology of our approach in a simplified setting, and to demonstrate that our approach reproduces the 
critical exponents obtained by conformal bootstrap \cite{Vasilev_nu_n2_1993,Gracey94,Iliesiu+18} and other methods \cite{Gat1990,Gracey_eta_n2_1991,Gracey_nu_n2_1992}.   

The bosonic propagator in anisotropic $d_L$-$d_Q$  nodal-point semimetals remains non-analytic, and therefore not perturbatively renormalizable, even near the upper critical 
dimension line $2d_L +d_Q=4$ \cite{sur_roy_semi_dirac}. This is a clear distinction from the case of relativistic fermions, with important consequences for the 
$\epsilon$ expansion. Approaching semi-Dirac fermions ($d_L=d_Q=1$) by expanding in the number of quadratic dimensions, $d_Q=2-\epsilon_Q$, $d_L=1$, one obtains leading 
corrections to critical exponents that are non-analytic and of the form $\sim\epsilon_Q\ln\epsilon_Q$ \cite{sur_roy_semi_dirac}. 
Here we show that the non-analytic dependence changes with the starting point on the upper critical line. Expanding in the number of linear dimensions, 
$d_Q=1$, $d_L=(3-\epsilon_L)/2$, we find leading $\sim\sqrt{\epsilon_L}$ corrections, putting the uniqueness of the $\epsilon$ expansion into question.

The outline of the paper is as follows: In Sec.~\ref{sec.GNY} we present the low energy theory of symmetry breaking of Dirac fermions, and derive the RG equations for the fermionic and bosonic velocities using a spherical and a cylindrical scheme in general dimension.
In Sec.~\ref{sec.COI} we utilize the soft cutoff RG formulation to determine the scaling of the bosonic propagator that results in cutoff independent corrections. We go on to calculate the GNY critical exponents to leading order in $1/N_f$. In Sec.~\ref{sec.SD} we extend the cutoff independence methodology to the family of anisotropic nodal-point semimetals, and calculate the exact $1/N_f$ critical exponents of the semimetal-to-insulator transition for the case of semi-Dirac fermions. Additionally, we present two separate $\epsilon$-expansions around the upper critical dimension line of anisotropic nodal-point semimetals. Finally, in Sec.~\ref{sec:discussion} we summarize the key results, compare them to the literature and discuss possible future research.

\section{Apparent violation of Lorentz invariance of Dirac fermions away from the upper critical dimension}
\label{sec.GNY}

Here we illustrate that away from the upper critical dimension, one-loop perturbative expansions of theories of interacting gapless fermions at quantum critical points are dependent 
on the RG cutoff scheme, even if they are controlled by large $N_f$ flavors of fermions. To do so, we consider the effective field theory of interacting Dirac fermions that describes, 
for example, the quantum criticality of Dirac semimetal-insulator transitions on the honeycomb lattice in $d=2$ spatial dimensions. Throughout this section we will make reference to this 
concrete example, which is situated far away from the upper critical spatial dimension $d_{\text{uc}} = 3$. 

\begin{figure}[t]
 \includegraphics[width=1.0\columnwidth]{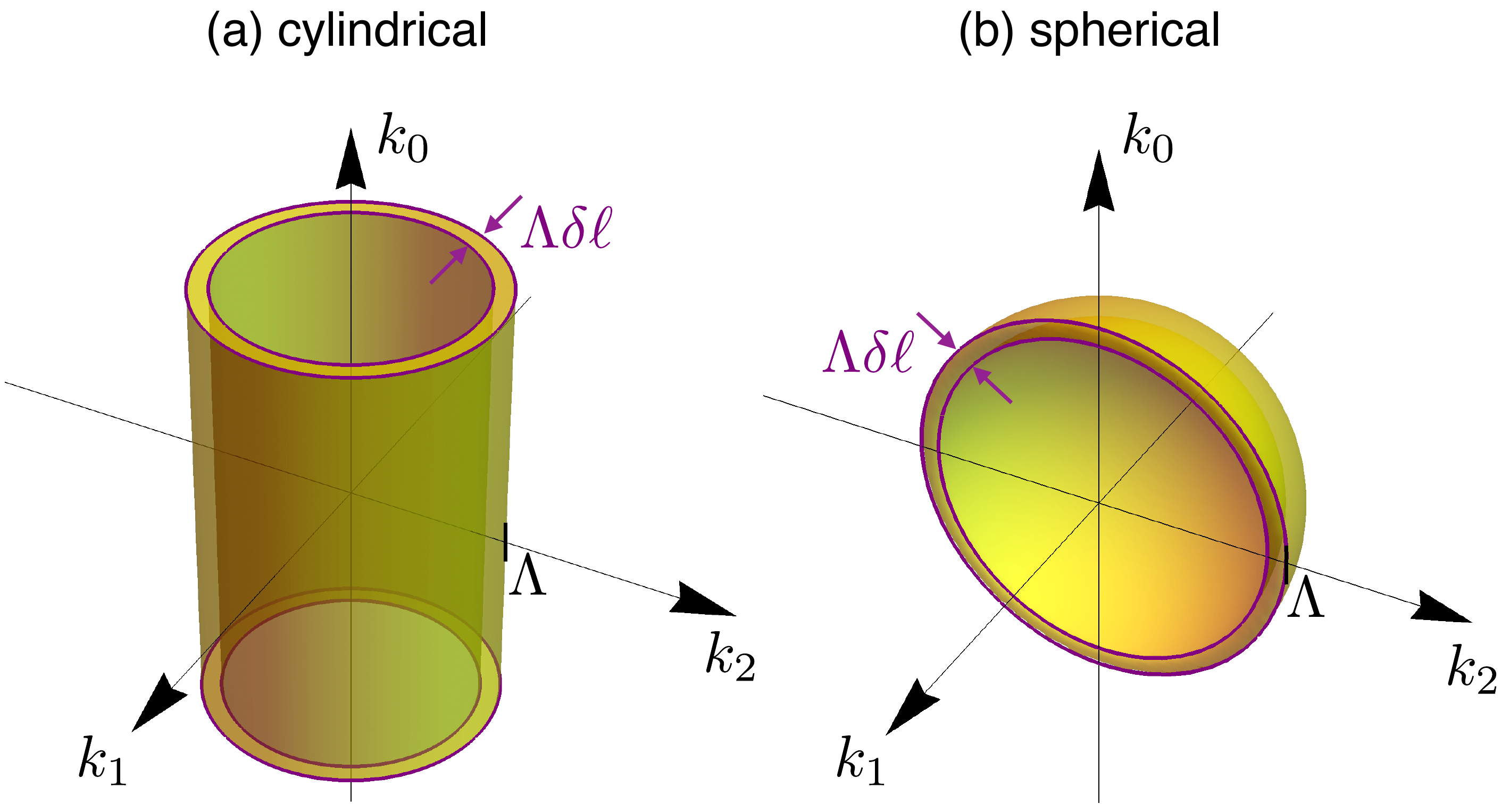}
 \caption{Wilson's infinitesimal shell RG integration schemes in $d=2$ spatial dimensions, at the cutoff scale $\Lambda$: (a) cylindrical and (b) spherical. Here $k_0$ and $\v{k}=(k_1,k_2)$
 denote frequency and momenta, respectively.}
\label{fig2}
\end{figure}

It is believed that for $1<d<3$ this quantum critical point possesses emergent Lorentz invariance, characterized by a dynamical exponent $z=1$ and a global terminal velocity, as has 
been observed with both the one-loop $\epsilon=3-d$ expansion of the effective Gross-Neveu-Yukawa (GNY) field theory \cite{herbut_roy_honeycomb} and lattice quantum Monte 
Carlo \cite{Assaad+13}. Here we analyze the GNY theory in $1<d<3$ dimensions, using Wilson's momentum-shell RG with two different cutoff schemes, 
shown in Fig.~\ref{fig2}.

Although the universal long-wavelength behavior should be independent of the choice of the cutoff scheme, we demonstrate that this is not the case within the perturbative momentum-shell 
framework. Using the cylindrical scheme in Fig.~\ref{fig2}(a), where the UV cutoff only acts on the spatial momentum directions and frequency is integrated over the whole real axis, 
we find an apparent violation of emergent Lorentz invariance. On the other hand, treating frequency and momenta on an equal footing and imposing an isotropic spherical cutoff in $D=d+1$ 
space-time dimensions, as shown in Fig.~\ref{fig2}(b),  we do find emergent Lorentz invariance. This contradiction is resolved in Section~\ref{sec.COI}.
 
\subsection{Gross-Neveu(-Yukawa) Models}

\noindent
The universality is captured by the Gross-Neveu (GN) model,
\begin{equation}
L_{\text{GN}} =  \bar{\Psi} ( \partial_\tau \gamma_0 + v \v{\partial} \cdot \v{\gamma} )\Psi + V( \bar{\Psi}\Psi )^2,
\label{gn_model}
\end{equation}
defined in imaginary time $\tau$ and $d$-spatial dimensions, such that $\v{\partial} = (\partial_1,\dots,\partial_d)$. The fermionic excitations possess a Fermi velocity $v$, and interact with the coupling strength $V$. We have generalized to $N_f$-component Dirac fields $\Psi=(\psi_1,\dots,\psi_{N_f})$, $\bar{\Psi}=\Psi^\dagger \gamma_0$. The Dirac $\gamma$ matrices anticommute $\{\gamma_\mu,\gamma_\nu\}=2 \delta_{\mu\nu}$ for $\mu,\nu = 0,\dots,d$, where $\v{\gamma} = (\gamma_1,\dots, \gamma_d)$ and the identity matrix is implicit. From this it follows that $\text{tr}\, \gamma_\mu \gamma_\nu = N_f \delta_{\mu\nu}$. Using this convention, the case of spinless fermions on the honeycomb lattice corresponds to $d=2$ and $N_f=4$, where it is customary to use the ``graphene representation'' \cite{herbut_graphene_2006}, $(\gamma_0,\gamma_1,\gamma_2) = (I_2 \otimes \sigma_z,\sigma_z \otimes \sigma_y, I_2 \otimes \sigma_x)$ with the Pauli matrices $\sigma_i$ and the $2\times2$ identity matrix $I_2$. 
We could then generalize to $N$ flavors of these 4-component fermions using $\gamma_\mu \rightarrow \gamma_\mu \otimes I_N$, such that $N_f = 4 N$.

We consider the case where strong interactions drive an instability in the $\bar{\Psi} \Psi$ channel at $V=V_c$. 
In the context of the honeycomb lattice, this corresponds to a quantum phase transition from the Dirac semimetal to a CDW insulator where the sublattice symmetry is 
spontaneously broken. More generally, we are studying the spontaneous symmetry breaking of a $Z_2$ Ising (pseudo)spin degree of freedom, which belongs to the chiral Ising GNY universality 
class \cite{Gross+74,ZinnJustin91}. 

It is natural to study this process by performing a Hubbard-Stratonovich transformation that redefines the problem as Dirac fermions coupled to an effective dynamical order parameter 
field $\phi$, conjugate to $\bar{\Psi}\Psi$.  This results in the Gross-Neveu-Yukawa (GNY) model,
\begin{align}
L_{\text{GNY}} &= \bar{\Psi}\left( \partial_\tau \gamma_0 + v \v{\partial} \cdot \v{\gamma}+  \frac{g}  {\sqrt{N_f}}  \phi\right) \Psi\notag \\
&\phantom{=}\,+ \frac{1}{2} \phi(-\partial_\tau^2-c^2\v{\partial}^2+ m^2) \phi+ \frac{ \lambda}{N_f} \phi^4.
\label{gny_model}
\end{align}

The Yukawa coupling anti-commutes with the non-interacting Hamiltonian and thereby fully gaps the fermionic quasiparticle spectrum upon condensation,
\begin{equation}
E(\v{k}) = \pm \sqrt{ v^2 \v{k}^2 + g^2\avg{\phi}^2 /N_f} .
\end{equation}

The order parameter mass $m^2$ is an RG relevant perturbation that tunes through the quantum critical point $m^2\sim V_c-V$. 
In contrast, although the Yukawa coupling $g$ and the self-interaction $\lambda$  are relevant at the non-interacting (Gaussian) fixed point, they are understood to flow to an infrared fixed point $(g_*,\lambda_*)$ in the critical plane $m^2=0$.

\subsection{Velocity RG equations in spherical and cylindrical RG schemes}

\begin{figure}
 \includegraphics[width=.95\columnwidth]{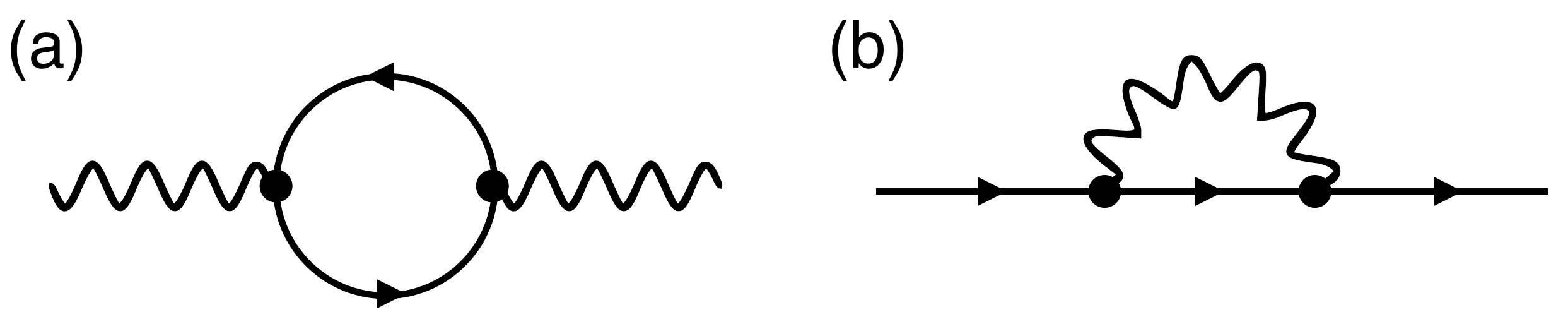}
\caption{One-loop self energy Feynman diagrams for the velocity RG equations. The fermion propagator is denoted by the arrowed line. The order parameter boson propagator is denoted by the wavy line.}
\label{fig3}
\end{figure}

For the purposes of the current discussion, it is sufficient to only study the flow of the Fermi velocity $v$ and the order parameter velocity $c$ in the vicinity of the GNY fixed point. 
We obtain the velocity RG equations at one-loop order, using Wilson's momentum-shell RG. In this approach, modes of highest energy near the ultraviolet cutoff scale $\Lambda$, corresponding 
to infinitesimal shells in Matsubara frequency $k_0$ and momentum $\v{k}=(k_1,\dots,k_d)$,  are integrated out.  
We consider the shell schemes displayed in Fig.~\ref{fig2},
\begin{align}
\text{(a) cylinder:}&\quad-\infty<k_0<\infty, \quad \Lambda e^{-\delta \ell} < \abs{\v{k}}<\Lambda,\\
\text{(b) sphere:}&\quad\Lambda e^{-\delta \ell} < \sqrt{k_0^2/v^2 + \v{k}^2}<\Lambda.
\end{align}
This is followed by the rescaling transformation
\begin{equation}
k_0 = k_0^\prime e^{-z \delta\ell},\quad \v{k} = \v{k}^\prime e^{-\delta\ell},
\end{equation}
where $z$ is the dynamical exponent. The quantum corrections are calculated from the one-loop fermion and boson self energy diagrams displayed in Fig.~\ref{fig3}.

Using the cylindrical cutoff scheme, the resulting RG equations for the velocities are given by
\begin{align}
\left(\frac{dv}{d\ell}\right)_{\text{cyl}} &= v \left(z-1 - g^2 \frac{2(v-c)-(d-3)c}{2N_fvc(v+c)^2}\right), \label{vcyl}\\
\left(\frac{dc}{d\ell}\right)_{\text{cyl}} &= c \left(z-1 - g^2 \frac{d(c^2-v^2)-(d-3)v^2}{ 16 dv^3 c^2} \label{ccyl}\right),
\end{align}
where we have made the rescaling $S_d\Lambda^{d-3} g^2 \rightarrow g^2$. Here $S_d$ denotes the surface area of the $d$-dimensional unit sphere,
\begin{equation}
S_d =\frac{1}{(2\pi)^d} \frac{ 2\pi^{\frac{d}{2}}}{\Gamma(d/2)}.
\end{equation}

On the other hand, using the spherical cutoff scheme in $D=d+1$ dimensions, we obtain the velocity RG equations  
\begin{align}
\left(\frac{dv}{d\ell}\right)_{\text{sph}} &= v \left(z-1 -  g^2 \frac{I_1\left(\frac{c}{v}\right) - I_0\left(\frac{c}{v}\right)}{N_fv^3}\right),\label{vsph}\\
\left(\frac{dc}{d\ell}\right)_{\text{sph}} &= c \left(z-1-g^2\frac{(D-2)(c^2-v^2)}{2 D v^3 c^2} \right),\label{csph}
\end{align}
where we have made the rescaling $S_D \Lambda^{D-4} g^2 \rightarrow g^2$ and defined the angular integrals $I_\mu(x)$ over the $D$-dimensional unit sphere 
$(\hat{k}_0^2 + \hat{\v{k}}^2 = 1)$,
\begin{equation}
I_\mu(x) = \frac{1}{S_D(2\pi)^D}\int d\hat{\Omega}  \frac{1-2\hat{k}^2_\mu}{\hat{k}_0^2 + x^2 \hat{\v{k}}^2}.
\end{equation}

From inspection of the (a) cylindrical (\ref{vcyl},~\ref{ccyl}) and (b) spherical (\ref{vsph},~\ref{csph}) RG equations in $d<3$, it is clear that $z=1$ and $c=v$ (for finite $g$) is not a fixed point solution for (a), but is a solution for (b). 
The putative emergent Lorentz invariance ($z=1$, $c=v$) is therefore violated for (a), but is satisfied for (b). 
This is the case even for $N_f\rightarrow \infty$ where the solution for (a) is $z=1$, $c=v(2-3/d)$. Naturally, each scheme will result in a different set of critical exponents.
However, for the $\epsilon=3-d$ expansion, where $g^2_* \propto \epsilon$, Lorentz invariance emerges for both shell schemes, which also share the same set of critical exponents.

This discussion demonstrates that away from the upper critical dimension, the perturbative loop expansion can lead to physically distinct conclusions at the same critical fixed point. 
Seemingly, the notion of universality breaks down, and the results depend on the way the cutoff RG scheme is implemented.
We resolve this apparent pathology in the next section, where we identify the conditions for quantum corrections that are independent of the RG scheme.

\section{Universal cutoff independent quantum corrections with soft cutoff RG and Landau damping: Dirac fermions}
\label{sec.COI}

Here we apply a completely general soft cutoff RG scheme to obtain the conditions for quantum corrections to be independent of the cutoff scheme, and therefore universal.
We prove that for interacting Dirac fermions, cutoff independent corrections are only obtained with a non-analytic order parameter propagator that scales as 
$k^{d-1}$ in $d$ spatial dimensions.

Remarkably, such dynamics arise from Landau damping by the gapless fermionic excitations, which is captured by the RPA resummation of fermion loop diagrams. 
Crucially, this non-perturbative effect is a product of the IR ($k\rightarrow0$) modes that are not typically accessible by perturbative means, such as the integration over infinitesimal shells.

With the soft cutoff RG procedure, we demonstrate the emergence of Lorentz invariance and calculate the GNY critical exponents in general dimensions to leading order in $N_f$. 
We find exact agreement with the conformal bootstrap results. 
Finally, we elucidate connections to the $\epsilon$-expansion.

\subsection{Soft cutoff RG scheme}

Following Refs.~\cite{brezin_zinn_justin_book,vojta_sachdev_dwave_field_theoretic_derivative,vojta_zhang_sachdev_d_wave_SC,Vojta+08,huh_sachdev_nematic_d_wave_SC}, we introduce the ultraviolet cutoff by means of a completely 
general, smooth, soft cutoff function, 
\begin{equation}
\label{eq.Alimits}
\begin{gathered}
A(z)  \sim \exp(-z^n) \quad(n>0),\\
A(z\rightarrow 0) = 1,\quad A(z\rightarrow\infty)=0.
\end{gathered}
\end{equation}

Within this description, the hard cutoff function is captured by $n\rightarrow \infty$. 
The soft cutoff procedure is implemented by augmenting the fermion and boson propagators with the cutoff function at the cutoff scale $\Lambda$,
\begin{equation}
G_{\Psi,\phi}(k) \rightarrow G_{\Psi,\phi}(k) A\left( \frac{a_\mu k_\mu^2}{\Lambda^2} \right) =  G_{\Psi,\phi}(k) A_k,
\end{equation}
where we define the $D=d+1$ dimensional $k_\mu=(k_0, \v{k})$ and use implicit summation over repeated $\mu,\nu=0,\dots,d$, such that $k^2=k_\mu k_\mu$. 
In the following we use the $A_k$ notation for brevity.

We explicitly include $a_\mu$ to make reference to the different cutoff schemes introduced in Sec.~\ref{sec.GNY}:
the cylindrical RG scheme corresponds to $a_0=0$, $a_{\mu\ne0} = 1$, whereas in the spherical scheme $a_0 = 1/v^2$, $a_{\mu\ne0} = 1$.
 
The quantum corrections to RG equations are then obtained by taking the logarithmic derivative in the cutoff scale $\Lambda \frac{d}{d\Lambda}$ of the one-particle irreducible 
vertex functions.  After all, RG is the re-summation of logarithmic divergences.
This is equivalent to the derivative $\frac{d}{d\ell}$ in the shell scheme with $\ell = \log (\Lambda/\Lambda_0)$.

\subsection{Cutoff scheme independence}
Whilst Dirac fermions are fundamental objects that propagate as
\begin{equation}
G_\Psi(k)=i \frac{k_0 \gamma_0 + v \v{k}\cdot \v{\gamma}}{k_0^2 + v^2 \v{k}^2},
\end{equation}
the \textit{effective} order parameter fields are not, and hence we should not necessarily expect them to adhere to the bare analytic dynamics of Eq.~\eqref{gny_model}.
Instead, we define a general homogeneous form of the boson propagator (at $m^2=0$),
\begin{equation}
G_\phi(k) = \frac{G_\phi(\hat{k})}{y^{n_\phi}},
\end{equation}
where $k = y \hat{k}$ with $\hat{k}^2=1$.

Now we determine the required scaling form of $G_\phi$ (through a constraint on $n_\phi$) to achieve cutoff scheme independent RG equations.
To do so, it is sufficient to calculate the quantum corrections from the one-loop fermion self energy diagram in Fig.~\ref{fig3}(b), 
\begin{equation}
\label{eq.fermion_self}
\frac{d}{d\ell} \Sigma(q) =  -\Lambda \frac{d}{d\Lambda} \frac{g^2}{N_f} \int_k G_\Psi (k+q)   A_{k} G_\phi(k)A_k,
\end{equation}
where $\int_k = \int d^Dk/(2\pi)^D$. Notice that the external $q$ dependence has been excluded from the cutoff function, and instead $A$ only regulates the internal $k$ integral. 
This is perfectly consistent with the conventional procedure in the hard cutoff RG. To further justify this step, we explicitly demonstrate in Appendix~\ref{app.soft1} that 
 linear $q$ contributions from $A_{k+q}$, that would contribute to the renormalization of the fermion propagator, do indeed vanish.

After expanding the right-hand side of Eq.~(\ref{eq.fermion_self}) to linear order in the external $q_\mu$, taking the logarithmic derivative and 
enacting the transformation $k = y \hat{k}= \tilde{y} \Lambda \hat{k}$, we obtain
\begin{equation}
\label{eq.fermion_self2}
\frac{d}{d\ell} \Sigma =  \frac{4i q_\mu \gamma_\mu }{\Lambda^{n_\phi +2 - D}}  \int_{\hat{\Omega} }\int_0^\infty \frac{\tilde{y}^{D}d\tilde{y}}{\tilde{y}^{n_\phi+1}} 
( 2 \hat{k}_\mu^2-1) a_\nu \hat{k}_\nu^2 \hat{G}_\phi A^\prime A, 
\end{equation}
where $A = A(\tilde{y}^2 a_\mu \hat{k}_\mu^2)$, $\hat{G}_\phi = G_\phi(\hat{k})$ and $\int_{\hat{\Omega}}$ is the $(D-1)$ dimensional angular integral scaled by $(2\pi)^D$. 
Note that in the above we have rescaled units such that $v=1$. 
It is simple to extract the cutoff independence by insisting that the result does not contain the UV cutoff scale $\Lambda$, which provides the constraint 
\begin{equation}
\label{eq.cutoffcond}
n_\phi = D-2 = d-1.
\end{equation}

Imposing this constraint, the radial integral can be evaluated with the substitution $u=A$ for general $A$. 
The result is now explicitly independent of $a_\mu$ (i.e. spherical or cylindrical schemes),
\begin{equation}
\label{eq.fermion_self3}
\frac{d}{d\ell} \Sigma =  i q_\mu \gamma_\mu  \int_{\hat{\Omega} } (2 \hat{k}_\mu^2-1) G_\phi(\hat{k}). 
\end{equation}

In contrast, if the constraint is not satisfied, the integral in Eq.~(\ref{eq.fermion_self2}) will depend on the explicit form of the cutoff function $A$, and cannot possibly contribute 
to universal phenomena. 

In fact, cutoff independence of a quantum correction that corresponds to a diagram with $N$ internal $D$-dimensional momenta can be determined by 
inspection of the scaling in the global radial coordinate $y$, $k_1=y  \hat{k}_1$, $k_2=y x_2  \hat{k}_2,\ldots k_N=y x_N  \hat{k}_N$. 
Such a quantum correction is independent of the cutoff scheme, and therefore universal, if the integrand scales as $1/y$. 
This follows from the identity
\begin{equation}
\Lambda \frac{d}{d\Lambda} \int_0^\infty \frac{dy}{y} \prod_i A^{n_i}\left(\frac{y^2 f_i}{\Lambda^2}\right) = 1, 
\label{identity}
\end{equation}
for positive integers $n_i$ and non-trivial angular functions $f_i=f_i(\hat{\Omega}_1,\ldots,\hat{\Omega}_N,x_2,\ldots x_N)$, which is proven in Appendix~\ref{app.soft2}. 

Finally, we should reiterate that \textit{any} RG scheme (Wilson's momentum shell, minimal subtraction, etc.) is applicable, provided the $n_\phi$ condition
that ensures cutoff independence is satisfied.

\subsection{Cutoff independent RPA propagator}

That $n_\phi=D-2$ should not be a surprise. This result is in agreement with the familiar form $G_\phi^{-1} \sim k$ in two spatial dimensions \cite{zinn_justin_book,
vozmediano_honeycomb_1994,vozmediano_graphene_columb1999,dtSon_coulomb_graphene}
and naturally arises at the $N_f\rightarrow\infty$ GNY fixed point, at which there is a large $\mathcal{O}(1)$ correction to the scaling of the order parameter field, $\eta_\phi =4-D$. 
This is a consequence of the one-loop fermion diagram in Fig.~\ref{fig3}(a), indicating that it is a Landau damping phenomena from the gapless fermionic excitations. 
Accounting for the anomalous scaling, the propagator $G_\phi^{-1} = k^{2-\eta_\phi}$ satisfies the condition for cutoff scheme independence. 

Away from the upper critical dimension, the cutoff independent propagator is non-analytic and so is not perturbatively renormalizable. 
This suggests that a non-perturbative solution is required. 

\begin{figure}[t]
 \includegraphics[width=.95\columnwidth]{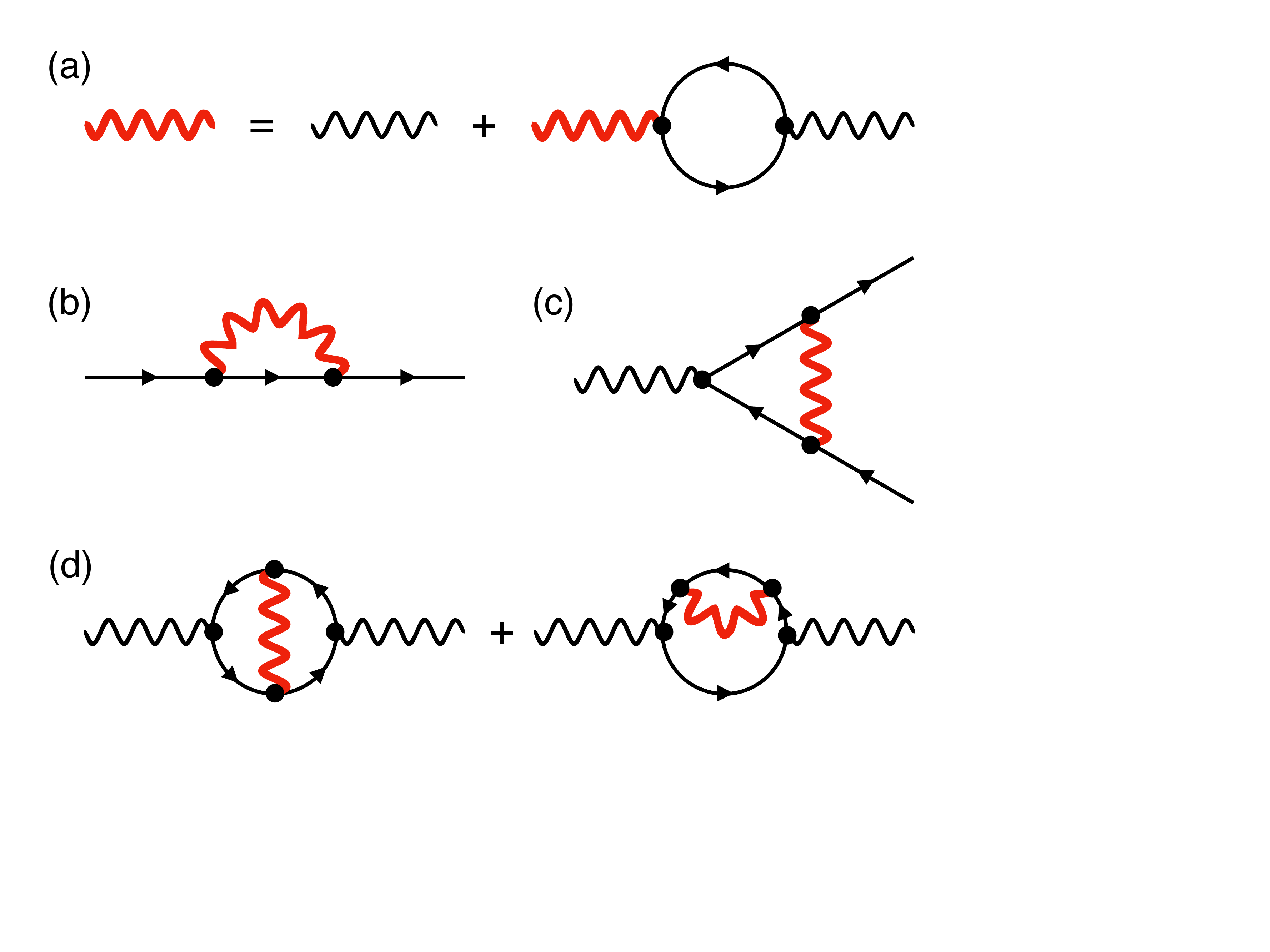}
 \caption{Feynman diagrams for large $N_f$ theories. (a) The bold way line represents the RPA boson propagator of the order parameter field. The fermion loops are integrated over the full range of modes and are self-consistently re-summed to infinite order. This results in a non-analytic Landau damped propagator that satisfies cutoff scheme independence. (b) The fermion self energy renormalizes the fermion propagator (arrowed straight line). (c) The vertex correction renormalizes the Yukawa coupling $g$. (d) The two loop diagrams renormalize the boson mass, and contribute to the correlation length exponent.}
\label{fig4}
\end{figure}

To self-consistently account for the damped boson dynamics, and to achieve cutoff independence, we use the non-perturbative RPA re-summation of fermion loops, which is shown 
diagrammatically in Fig.~\ref{fig4}(a), to obtain the dressed boson propagator
\begin{equation}
G_\phi^{-1}(q) = G_{\phi,0}^{-1}(q) + \Pi(q), 
\end{equation}
where $G_{\phi,0}^{-1}(q)=q_0^2+c^2\v{q}^2+ m^2$ is the bare boson propagator. The bosonic self energy 
\begin{equation}
\Pi(q) = \frac{g^2}{N_f} \int_k \text{tr}\,G_\Psi(k+q) G_\Psi(k),
\end{equation}
is calculated by integrating over the full range of modes. Crucially, $\Pi$ includes the IR ($k\rightarrow 0$) modes that are not accounted for in Wilsonian shell schemes, and results in 
the non-analytic propagator (see Appendix~\ref{app.GNY1})
\begin{equation}
G_\phi^{-1}(k) =  \frac{g^2S_D \alpha_D }{v^{D-1}} (k_0^2 + v^2 \v{k}^2)^{\frac{D-2}{2}} + m^2,
\label{gny_RPA_propagator}
\end{equation}
where 
\begin{equation}
\label{alphaconst}
\alpha_D=   -\frac{\pi}{2\sin(\frac{\pi D}{2})} \frac{ \Gamma(D/2)^2} { \Gamma(D-1)}.
\end{equation}

Note that the IR scaling of the dressed RPA boson propagator (\ref{gny_RPA_propagator}) satisfies the condition of cutoff independence, Eq.~(\ref{eq.cutoffcond}).
We have neglected the sub-leading momentum terms in $G_{\phi,0}^{-1}(k)$ since these terms are irrelevant in an RG sense.
Formally, the RPA contribution dominates in the large $N_f$ limit, which is evident after making the rescaling $g^2 \rightarrow g^2 N_f$.

The Landau damped dynamics affects the scaling of the effective order parameter field. Crucially,  the quartic self interaction $\lambda \phi^4$ of Eq.~\eqref{gny_model} is rendered irrelevant at tree-level, and so is neglected in the following. This is a common feature of the ``interaction driven scaling'' \cite{ssLee_fermi_surf_2+1} of gapless fermionic systems.

\subsection{Large $N_f$ RG equations}

We now perform an RG analysis of the large $N_f$ field theory
\begin{equation}
\label{eq.GNYlarge_N}
L  = \bar{\Psi}\left( \partial_\tau \gamma_0 + v \v{\partial} \cdot \v{\gamma}+  \frac{g}  {\sqrt{N_f}}  \phi\right) \Psi + \frac{1}{2} \phi\, G_\phi^{-1} \phi,
\end{equation}
using the soft cutoff scheme to calculate the diagrams in Figs.~\ref{fig4}(b)-(d) to leading order in $N_f$.
Here $G_\phi(k)$ (\ref{gny_RPA_propagator}) is the fully dressed bosonic propagator that is obtained by the RPA re-summation depicted in Fig.~\ref{fig4}(a). 
As demonstrated in Sec.~\ref{sec.COI}, $G_\phi(k)$ has the correct 
IR scaling that ensures cutoff independence. This makes the evaluation of radial integrals trivial since we can simply use the radial integral identity, Eq.~(\ref{identity}). 
The remaining angular integrals of the 
one loop diagrams, such as Eq.~(\ref{eq.fermion_self3}) in the case of the fermion self-energy correction, are elementary and can be 
carried out analytically. Here we only present the results. Details of the calculation, e.g. on the evaluation of the angular integrals in general dimension, can be found in 
Appendix~\ref{app.GNY2}. 

\noindent
For the fermionic self-energy diagram, Fig.~\ref{fig4}(b), we obtain
\begin{eqnarray}
\frac{d}{d\ell} \Sigma(q) & = &  - \Lambda \frac{d}{d\Lambda} \frac{g^2}{N_f} \int_k G_\Psi (k+q)   G_\phi(k) A_k^2\nonumber \\
& = &   -i\frac{D-2}{\alpha_D D N_f}(q_0 \gamma_0 + v \v{q} \cdot \v{\gamma}),
\end{eqnarray}
with $\alpha_D$ defined in Eq.~(\ref{alphaconst}). The vertex correction, which is show in Fig.~\ref{fig4}(c) and which renormalizes the Yukawa coupling $g$, is equal to 
\begin{eqnarray}
\frac{d}{d\ell} \Xi & = & \Lambda \frac{d}{d\Lambda} \frac{g^3}{\sqrt{N_f}^3} \int_k     G^2_\Psi(k) G_\phi(k) A_k^3 \nonumber\\
& & = -\frac{1}{\alpha_DN_f}\frac{g}{\sqrt{N_f}}.
\end{eqnarray}

We further evaluate the two-loop diagrams in Fig.~\ref{fig4}(d) for zero external momentum and frequency but finite boson mass $m^2\neq 0$, since they contain 
quantum corrections of order $1/N_f$ that renormalize $m^2$,
\begin{eqnarray}
\frac{d}{d\ell} \tilde{\Pi} & = &  \Lambda \frac{d}{d\Lambda}\frac{g^4}{N_f^2} \int_{k,q} G_\phi(q) A_q \times \text{tr}\, \Big[  \nonumber \\
& & G_\Psi(k+q) G_\Psi(k+q) G_\Psi(k)G_\Psi(k)A^2_{k+q} A^2_k \nonumber\\
& & +2G_\Psi(k+q) G_\Psi(k) G_\Psi(k)G_\Psi(k)A_{k+q} A^3_k\Big] \nonumber\\
& = & \frac{D-1}{\alpha_D^2 N_f \sin(\frac{\pi D}{2})} \frac{\pi \Gamma(\frac{D}{2})^2}{\Gamma(D-1)}m^2.
\end{eqnarray} 
Details of the calculation of the two-loop diagrams can be found in Appendix~\ref{app.GNY3}. 

Under the RG transformation momentum and frequency are rescaled as
\begin{equation}
\v{k} = \v{k}' e^{-\delta\ell},\quad k_0=k_0' e^{-z\delta\ell},
\end{equation}
with $z$ the dynamical critical exponent. The fields are rescaled as
\begin{equation}
\Psi(k) = \Psi^\prime(k^\prime) e^{- \Delta_\Psi \ell /2},\quad \phi(k) = \phi^\prime(k^\prime) e^{- \Delta_\phi  \ell /2},
\end{equation}
where $\Delta_X = [X^\dagger X] +\eta_X$ ($X=\Psi,\phi$) are the sum of tree-level $[\ldots]$ and anomalous $\eta_X$ scaling dimensions.  The RG equations 
are obtained by combining quantum corrections and rescaling contributions,
\begin{eqnarray}
\label{rg_v}
\frac{dv}{d\ell} & = &  v \Big[-(D+z +\Delta_\Psi )+ \frac{D-2}{\alpha_D D N_f}\Big],\\
\frac{dg}{d\ell} & = & g\Big[ 2(1-D-z) - \Delta_\Psi - \frac{\Delta_\phi}{2} -\frac{1}{\alpha_D N_f} \Big],\\
\label{rg_m2}
\frac{dm^2}{d\ell} & = & m^2 \Big[-(D-1+z+   \Delta_\phi) \nonumber\\
& & \quad+ \frac{D-1}{\alpha_D^2 N_f \sin(\frac{\pi D}{2})} \frac{\pi \Gamma(\frac{D}{2})^2}{\Gamma(D-1)}\Big],
\end{eqnarray}
subject to the constraint 
\begin{equation}
\Delta_\Psi = 1-D-2z+ \frac{D-2}{\alpha_D D N_f},
\end{equation}
which follows form the scale invariance of $\int \bar{\Psi}k_0 \gamma_0\Psi$.

The solution $z=1$ of Eq.~(\ref{rg_v}), for all $v$, indicates the emergence of Lorentz invariance at the quantum critical point.
Moreover,  $g$ is scale invariant since it can be scaled out of the large $N_f$ field theory (\ref{eq.GNYlarge_N}), using $\phi\rightarrow \phi/g$, 
$m^2 \rightarrow g^2 m^2$. From this it follows that 
\begin{equation}
\Delta_\phi = 2-2D-4 \frac{D-1}{\alpha_D D N_f}.
\end{equation}

Effectively, Eq.~(\ref{eq.GNYlarge_N}) describes the $(g_*,\lambda_*)$ GNY critical fixed point of Eq.~\eqref{gny_model}, at which $g$ and $\lambda$ are 
irrelevant perturbations. The correlation length exponent $\nu$ is determined by the flow of the single relevant perturbation at the critical fixed point,
\begin{equation}
\frac{dm^2}{d \ell} = \nu^{-1} m^2,
\end{equation}
and can therefore be extracted from Eq.~(\ref{rg_m2}). The resulting critical exponents in $D=d+1$ dimensions, to leading order in $N_f$, are 
\begin{eqnarray}
\label{gny_exponents1}
\eta_\Psi & = &  \frac{2(2-D)}{DN_f}  \frac{\sin(\frac{\pi D}{2})}{\pi} \frac{\Gamma(D-1)}{\Gamma(\frac{D}{2})^2},\\
\label{gny_exponents2}
\eta_\phi & = &  \frac{8}{D N_f}\frac{\sin(\frac{\pi D}{2})}{\pi} \frac{\Gamma(D)}{\Gamma(\frac{D}{2})^2},\\
\label{gny_exponents3}
\nu^{-1} & = &  D-2 + \frac{D-2}{ N_f}\frac{\sin(\frac{\pi D}{2})}{\pi} \frac{\Gamma(D+1)}{\Gamma(\frac{D+2}{2})^2},
\end{eqnarray}
which have been extracted using $[\bar{\Psi} \Psi] = -(D+1)$ and $[\phi\phi] = -2(D-1)$. 
These exponents are in agreement with previous results using the large $N_f$ conformal bootstrap \cite{Vasilev_nu_n2_1993,Gracey94,Iliesiu+18} and the critical point 
large $N_f$ formalism \cite{Gat1990,Gracey_eta_n2_1991,Gracey_nu_n2_1992}.

\subsection{Connections to the $\epsilon$-expansion}

Finally we draw some connections to the $\epsilon=4-D$ expansion below the upper critical dimension. To zeroth order in $\epsilon$, the bare bosonic 
propagator $G_\phi^{-1}(k)\sim k^2$ satisfies the condition (\ref{eq.cutoffcond}) for cutoff independence, $n_\phi=2-\epsilon$.
The $\mathcal{O}(\epsilon)$ corrections to perturbative loop diagrams are cutoff dependent, but do not enter RG equations at any order in $\epsilon$.
 The GNY RG equations from the $\epsilon$-expansion are therefore independent of the cutoff scheme. As we will see in Sec.~\ref{sec.SD}, this is the crucial difference 
to the case of anisotropic nodal fermions, where the bare propagator results in cutoff dependent contributions even near the line of upper critical dimensions.

Naturally, for $D=4-\epsilon$ the critical exponents $\eta_\Psi$ (\ref{gny_exponents1}), $\eta_\phi$ (\ref{gny_exponents2}) and $\nu$ (\ref{gny_exponents3}) agree to leading 
order in $N_f$ with those obtained from the $\epsilon$-expansion, order by order in $\epsilon$.  
The $\epsilon$-expansion, however, can also be formulated using the scheme outlined in Fig.~\ref{fig4}, and results in the RG equations already evaluated at the $(g_*,\lambda_*)$ 
critical fixed point. 

As a first step,  the $1/\epsilon$ pole of the RPA propagator (\ref{gny_RPA_propagator}) must be extracted. The prefactor to the pole can also be obtained from the logarithmic 
divergence of $\Pi$ in $D=4$.  Then the remaining diagrams are evaluated using 
\begin{equation}
G_\phi^{-1}(k) = \frac{1}{\epsilon} \frac{g^2}{16 \pi^2} (k_0^2 + v^2 \v{k}^2) + m^2.
\end{equation}

It can be verified that the quantum loop corrections calculated in this manner agree with those obtained by perturbative means, after solving for the fixed 
point $(g_*,\lambda_*)\sim\mathcal{O}(\epsilon)$. 

Although somewhat trivial for GNY theories, this methodology can act as an independent check of large $N_f$ critical exponents obtained from the $\epsilon$ expansion. 
It has cutoff scheme independence encoded through the RPA propagator, which is crucial when considering anisotropic systems such as those considered in the next section. 
Finally, it is a shortcut to accessing the critical fixed point, which is valuable when dealing with a complicated set of RG equations.

\section{General $d_L$ and $d_Q$ system}
\label{sec.SD}

In this section we employ the soft cutoff approach to determine the universal critical behavior of nodal-point semimetals with $d_L$ linear and $d_Q$ quadratic momentum directions. 
Quantum phase transitions between anisotropic nodal point semimetals and insulating ordered states can be driven by generic short range interactions between the fermionic
quasiparticles. An overview of various interaction-driven instabilities of semi-Dirac fermions and their competition as a function of different microscopic interaction parameters can be found in 
Ref.~\cite{Roy+18}.  As in the case of relativistic Dirac fermions, we consider a scalar order parameter field $\phi$, for simplicity.  Such an Ising order parameter field could for 
example describe CDW order.

We compute the exact critical exponents for semi-Dirac fermions ($d_L=d_Q=1$) to leading order in $1/N_f$. Finally,  from our results for general $d_L$, $d_Q$ we obtain $\epsilon$ expansions 
in both the number of linear and quadratic dimensions.

\subsection{Effective Field Theory}

The effective Yukawa theory for anisotropic nodal-point fermions with $d_L$ linear and $d_Q$ quadratic momentum directions in $d=d_L+d_Q$ spatial dimensions, coupled to the 
scalar bosonic order parameter field $\phi$ is given by
\begin{align}
\label{eq.L_dL_dQ}
L &= \bar{\Psi} [\partial_\tau \gamma_0 + \v{\partial}_L \cdot \v{\gamma}_L + (i v_Q^2 \v{\partial}_Q^2+\Delta) \gamma_Q]\Psi\notag\\
&\phantom{=}\,+ \frac{g}{\sqrt{N_f}} \phi \bar{\Psi}\Psi + \frac{1}{2} \phi \,G^{-1}_{\phi} \phi,
\end{align} 
where we have generalized to $N_f$ copies of fermions, $\Psi = (\psi_1,\cdots,\psi_{N_f})$, $\bar{\Psi} = \Psi^\dagger \gamma_0$, and defined 
$\v{\partial}_L = (\partial_1,\dots,\partial_{d_L})$ and $\v{\partial}_Q = (\partial_{d_L+1},\dots,\partial_{d_L+d_Q})$. The parameter $v_Q$ is related 
to the curvature of the quadratic dispersion. The linear momenta couple to $\v{\gamma}_L = (\gamma_1,\dots,\gamma_{d_L})$, which together with $\gamma_0$ and $\gamma_Q$ 
form a set of $d_L+2$ mutually anti-commuting gamma matrices, $\{\gamma_\mu,\gamma_\nu\}=2\delta_{\mu\nu}$.
Note that for $d_Q=0$ the model reduces to the large-$N_f$ GNY theory, defined in Eq.~(\ref{eq.GNYlarge_N}).

The Yukawa coupling anti-commutes with the non-interacting Hamiltonian and thereby fully gaps the fermionic quasiparticle spectrum upon condensation of
the order parameter, 
\begin{equation}
\label{eq.disp_ani}
E(\v{k})= \sqrt{\v{k}_L^2 + (v_Q^2 \v{k}_Q^2+\Delta)^2 + g^2\avg{\phi}^2/N_f },
\end{equation}
where $\v{k}_L = (k_1,\dots,k_L)$, $\v{k}_Q = (k_{d_L+1},\dots,k_{d_L+d_Q})$, and $\v{k}=(\v{k}_L,\v{k}_Q)$. 

The parameter $\Delta$  tunes the system through a topological phase transition from a nodal-surface semimetal ($\Delta<0$) to a 
trivial band insulator ($\Delta>0$). The nodes for $\Delta<0$ are given by the $d_Q$ dimensional sphere $\v{k}_Q^2=-\Delta/v_Q^2$ for $\v{k}_L=0$. 
The experimentally most relevant cases are nodal line semimetals for $d_Q=2$ and semimetals with a pair of isolated Weyl points for $d_Q=1$.
Both $\Delta$ and the order parameter mass $m^2$ are relevant perturbations at the multi-critical point, $\Delta=0$ and $m^2=0$. 

Since all quadratic directions couple to the same matrix $\gamma_Q$ the dispersion remains radially symmetric in the $d_Q$ subspace. A different class of 
semimetals can be defined in terms of spherical harmonics that couple to different $\gamma$ matrices \cite{herbut_janssen_qbt_top_mott_QBT,janssen_herbut_qbt_nematic}. 
Such theories, which could describe rotational symmetry breaking (nematic transitions) in the $d_Q$ subspace, are not considered here. 

\noindent
The dressed RPA boson propagator is given by 
\begin{eqnarray}
G_{\phi}^{-1}(k) & = & G_{\phi,0}^{-1}(k) + \Pi(k) \nonumber\\
& = & c_L\left(k_0^2 + \v{k}_L^2\right) + c_Q \v{k}_Q^2 + m^2+ \Pi(k),
\end{eqnarray}
where $k=(k_0,\v{k})$.  As we will see later, the bosonic self energy correction $\Pi(k)$, which arises from the damping of the order-parameter fluctuations by 
the anisotropic nodal fermions, dominates the long-wavelength behavior of $G_{\phi}(k)$. As in the case of the GNY theory, the resulting IR scaling of $G_{\phi}(k)$
satisfies the condition of cutoff independence.

Note that the bare boson propagator $G_{\phi,0}^{-1}(k)$ does not inherit the unusual anisotropy of the fermionic quasiparticle dispersion $E(\v{k})$ (\ref{eq.disp_ani}) 
but instead depends quadratically on both $\v{k}_L$ and $\v{k}_Q$. This conventional form arises naturally from integrating out fermion modes near the UV cutoff in 
perturbative RG schemes.

\subsection{Scaling}

We start by a scaling analysis of the Yukawa-type field theory for $d_L$-$d_Q$ fermions, given in Eq.~(\ref{eq.L_dL_dQ}). 
To account for the different scaling of linear and quadratic momenta, we define \emph{two} 
scaling exponents, $z_L$ and $z_Q$, such that 
\begin{equation}
\label{eq.dLdQresc1}
k_0 = k_0^{\prime} e^{-z_L \delta\ell},\ \v{k}_L = \v{k}_L^{\prime} e^{-z_L \delta\ell},\ \v{k}_Q = \v{k}_Q^{\prime} e^{-z_Q \delta\ell},
\end{equation}
under rescaling. This allows us to unify the different scaling conventions in previous studies of anisotropic systems: ($z_L = 2$, $z_Q=1$)~\cite{nagaosa_semi_dirac_nature, Cho+16, uryszek_semi_dirac, sur_roy_semi_dirac} and ($z_L = 1$, $z_Q = 1/2$)~\cite{nagaosa_semi_dirac_prl,Wang+17,Roy+18}.

 Note that we have rescaled frequency and linear momenta with the same exponent. In general, one should consider different exponents 
$z_0$ and $z_L$ and allow for a renormalization of the Fermi velocity $v_L$ along the linear momentum direction. However, exactly as for the purely relativistic case, there is 
an emergent Lorentz invariance in the $k_0$-$\v{k}_L$ subspace at the critical fixed point. For that reason we have set $z_0=z_L$ and $v_L=1$, without loss of generality.  

At tree-level, the gapless fermionic quasiparticle energy $E(\v{k})$ at the multi-critical point ($\Delta=0$, $m^2=0$, $\avg{\phi}=0$) scales as
\begin{equation}
E(\v{k}) = E(\v{k}^\prime) e^{-z_L\delta\ell},
\end{equation}
under the condition that $[v_Q^2]+2z_Q = z_L$. The latter is satisfied if $v_Q$ is scale invariant, $[v_Q]=0$, and $z_Q/z_L =1/2$.
The RG procedure could therefore be established by integrating out modes from the $D=d_L+d_Q+1$ dimensional shell
\begin{equation}
\label{eq.dLdQshell}
\Lambda e^{-z_L \delta \ell} \le \sqrt{k_0^2 + E^2(\v{k})} \le \Lambda
\end{equation}
below the UV cutoff $\Lambda$. The additional factor of $z_L$ in the exponent suggests that one should consider $z_L \delta \ell$ as the ``unit length'' and define the rescaling
of the fields as
\begin{align}
\label{eq.dLdQresc2}
\Psi(k) &= \Psi^{\prime}(k^{\prime})e^{-\Delta_{\Psi}z_L\delta\ell/2}\\
\label{eq.dLdQresc3}
\phi(k)&=\phi^{\prime}(k^{\prime})e^{-\Delta_{\phi}z_L\delta\ell/2}
\end{align}
where $\Delta_\Psi = [\bar{\Psi}\Psi]+\eta_\Psi$ and $\Delta_\phi = [\phi\phi]+\eta_\phi$ are the critical dimensions of 
the fermionic and bosonic fields, respectively. With these conventions the universal critical behavior will only depend on the ratio $z_Q/z_L$. 
Scale invariance of the free-fermion action at tree-level requires that 
\begin{align}
[\bar{\Psi}\Psi] &= -\left(2 + d_L+d_Q/2\right).
\label{eq.fermion_TL}
\end{align}

\begin{figure}[t!]
 \includegraphics[width=\columnwidth ]{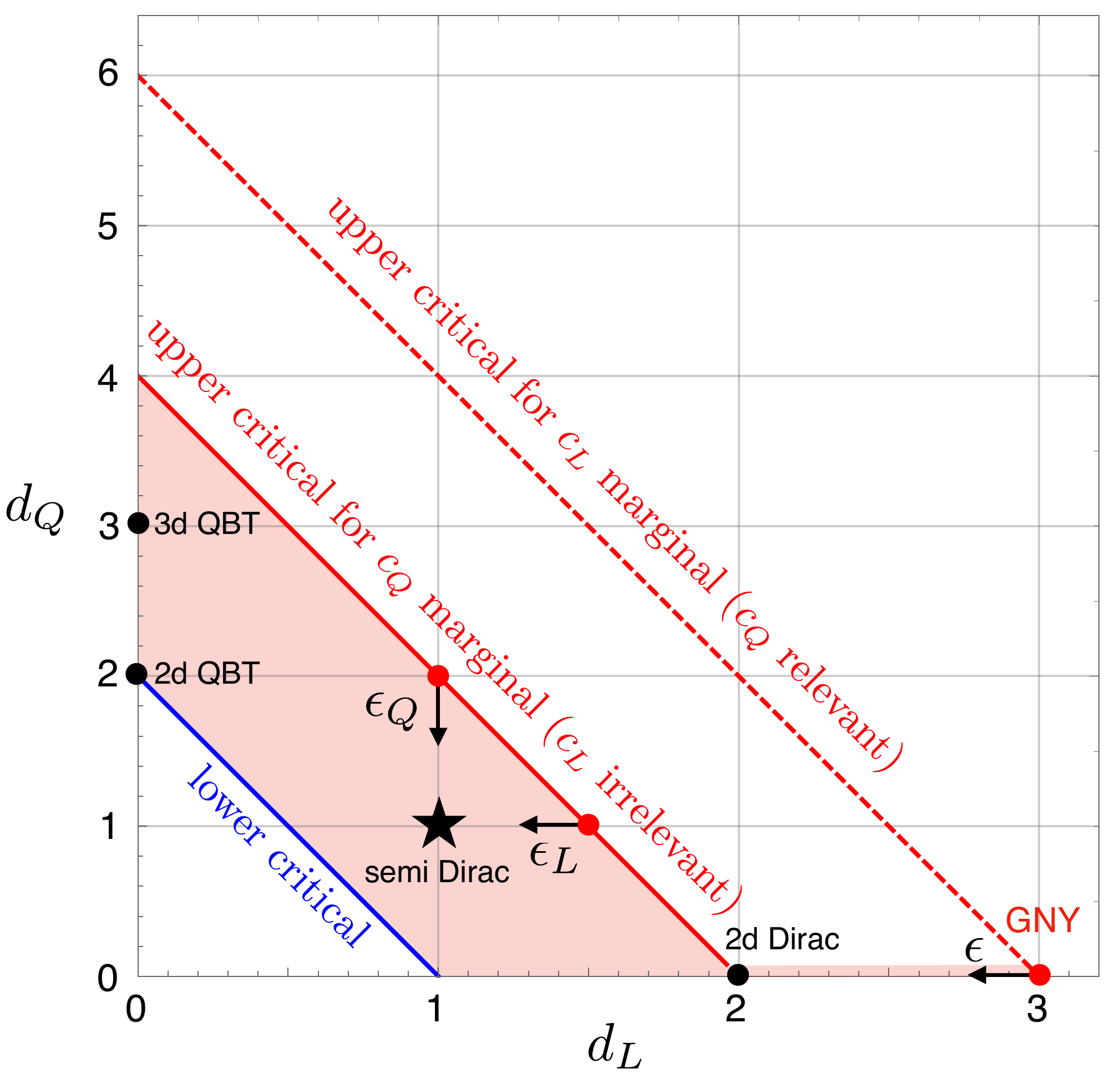}
 \caption{Lower and upper critical dimension lines of nodal point semimetals with $d_L$ linear and $d_Q$ quadratic momentum directions. 
 For $d_Q>0$ the line of upper critical dimensions $2d_L+d_Q=4$ (red solid line) is obtained from the condition that $c_Q$ is scale invariant. 
 The dashed red line is obtained from the condition that $c_L$ is scale invariant and therefore terminates at the upper critical dimension 
 $d_L=d_\textrm{uc}=3$ of the GNY theory. The universal critical behavior of semi-Dirac fermions ($d_L=d_Q=1$) could be approached by $\epsilon$
 expansions in both the number of linear and quadratic dimensions.}
\label{fig5}
\end{figure}

We now turn our attention to the bosonic sector. Since the bare order parameter propagator 
\begin{equation}
\label{eq.bareboson}
G_{\phi,0}^{-1}(k) = c_L\left(k_0^2 + \v{k}_L^2\right) + c_Q \v{k}_Q^2,
\end{equation}
does not show the same anisotropic momentum scaling as the fermionic quasiparticles, but instead depends quadratically on both $\v{k}_L$ and $\v{k}_Q$, scale invariance is violated at the bare, 
non-interacting level. 

For systems with a finite number of quadratic band touching directions, $d_Q>0$, it is natural to choose the boson scaling 
\begin{equation}
\label{eq.boson_TL}
[\phi\phi] = -  (2+d_L + d_Q/2),
\end{equation}
such that $c_Q$ is marginal, but $c_L$ is irrelevant. The resulting tree-level scaling dimension of the Yukawa coupling 
is given by
\begin{equation}
[g] = \frac{1}{4} (4- 2d_L - d_Q),
\end{equation}
defining an upper critical line $2 d_L+d_Q=4$ of marginal interactions, shown in Fig.~\ref{fig5}. Note that this line does not contain the upper-critical dimension 
$d_L=d_\textrm{uc}=3$ of the GNY theory ($d_Q=0$).  This point is the termination of the upper critical line $2 d_L+d_Q=6$ obtained from $[c_L]=0$ and $[g]=0$. 
Note that in this case $c_Q$ relevant.

\subsection{Cutoff independence and dressed RPA boson propagator}\label{subsec:dL_dQ_RPA}

The lack of scale invariance of the bare bosonic order parameter propagator $G_{\phi,0}^{-1}(k)$ (\ref{eq.bareboson}) in $d_L$-$d_Q$ fermion systems 
is intimately linked to the fundamental problem that perturbative RG procedures do not correctly account for the long-wavelength fluctuations of the order parameter, e.g. they neglect
 the phenomenon of Landau damping. 

The irrelevance of $c_L$ suggests an IR divergence on approach to the critical fixed point. Previously it was argued that this unphysical divergence should be 
regulated with the asymptotic self-energy correction $\Pi(k_0,\v{k}_L,\v{k}_Q=0)$ along the linear momentum and frequency directions  \cite{sur_roy_semi_dirac,uryszek_semi_dirac}. 

Within our soft cutoff approach it is clear, however, that below the upper critical line such a partially dressed boson propagator leads to quantum corrections that are dependent 
upon the the UV cutoff scheme and hence non-universal. By enforcing that the results are independent of the cutoff scheme we can 
deduce the correct IR scaling of the dressed boson propagator.  

Since the dressed boson propagator should inherit the different scaling of momenta along linear and quadratic directions, we make the \textit{Ansatz}
\begin{equation}
\label{eq.bosonpropscaling}
G_{\phi}(k) = \frac{G_{\phi}(\hat{k})}{\varepsilon^{n_{\phi}}},
\end{equation}
where we have defined the $(d_L+2)$ dimensional vector 
\begin{equation}
\label{eqn:SD-varepsilon}
\varepsilon_\mu = (k_0,\dots,k_{d_L}, \v{k}_Q^2),
\end{equation}
and $\varepsilon^2 = \varepsilon_\mu\varepsilon_\mu$, using implicit summation over $\mu$. 

In the soft cutoff approach we dress boson and fermion propagators with a completely general cutoff function $A$,
\begin{equation}
G_{\Psi,\phi}(k) \rightarrow G_{\Psi,\phi}(k) A\left( \frac{a_\mu\varepsilon_\mu^2}{\Lambda^2} \right),
\end{equation}
which only needs to satisfy the boundary conditions $A(0)=1$ and $\lim_{z\to\infty}A(z) = 0$. The hard cutoff 
is included as the special case where $A$ is a step function, $A(z) = \Theta(1-z)$. We can also include coefficients $a_\mu$ 
to allow for different cutoff schemes, e.g. $a_\mu=1$ for $\mu=0,\cdots,d_L+1$ corresponds to the spherical 
scheme of Eq.~(\ref{eq.dLdQshell}), while in the cylindrical scheme, $a_0=0$, $a_{\mu\neq0}=1$, the cutoff only 
acts on the spatial momenta.

Cutoff independence means that the quantum corrections do not depend upon the the UV scale $\Lambda$, the 
cutoff function $A$, and the choice of coefficients $a_\mu$. As discussed in detail in Sec.~\ref{sec.COI}, this is the case if the 
integrands of the loop corrections scale as $1/\varepsilon$, since all cutoff dependence vanishes
due to the radial integral identity (\ref{identity}) for $y=\varepsilon$. For the $d_L$-$d_Q$ system this is only the case 
if the dressed boson propagators scales with the exponent
\begin{equation}
\label{eq.cutoffinddLdQ}
n_{\phi} = d_L + d_Q/2 -1.
\end{equation}

As in the case of relativistic Weyl or Dirac fermions, the fully dressed RPA boson 
propagator $G_{\phi}^{-1}(k) =  G_{\phi,0}^{-1}(k) + \Pi(k)$ satisfies the condition (\ref{eq.cutoffinddLdQ}) of cutoff 
independence in the long-wavelength limit.  It is not possible, however,  to obtain a closed expression for the bosonic self energy $\Pi(k)$. 
The asymptotic forms of $\Pi(k)$ for $2<2d_L+d_Q\le4$ and $d_Q>0$ along the linear and quadratic directions is given by 
\begin{equation}\label{eqn:dL-d_Q_RPA_asymptotic_form}
\Pi(k) \sim \left\{ \begin{array}{cc}  (k_0^2+  \v{k}_L^2)^{\frac{1}{4} (2d_L+d_Q-2)} & \textrm{for }    \v{k}_Q=0\\
  \v{k}_Q^{(2d_L+d_Q-2)}  & \textrm{for } k_0,\v{k}_L = 0   \end{array}  \right.
\end{equation}

Below the upper critical line $2d_L+d_Q=4$, the self energy $\Pi(k)$ dominates over the bare terms in the propagator  in the $k\to0$ limit
and therefore determines the universal critical behavior. The resulting propagator  $G_{\phi}^{-1}(k)$ is inherently anisotropic, reflecting 
the different scaling of momenta, and non-analytic, showing that is is inaccessible by perturbative means. It strongly scales with the 
dimensions $d_L$, $d_Q$ of the system, $G_{\phi}^{-1}\sim \varepsilon^{d_L + d_Q/2 -1}$, satisfying the condition (\ref{eq.cutoffinddLdQ}) and 
resulting in cutoff independent quantum corrections. 

Note that the boson propagator $G_{\phi}^{-1}$ in anisotropic nodal fermion systems remains non-analytic even along the line 
of upper critical dimensions. Although in this case the conventional scaling $\sim \v{k}_Q^2$ along the quadratic directions is recovered, 
the IR scaling along the linear directions, $G_{\phi}^{-1} \sim\sqrt{k_0^2 + \v{k}_L^2}$ remains non-analytic. 
As demonstrated in Ref.~\cite{sur_roy_semi_dirac} and investigated in more detail in Sec.~\ref{sec.epsilon}, this has important consequences for $\epsilon$ 
expansions below the line of upper critical dimensions.

\subsection{Large $N_f$ RG equations for general $d_L$, $d_Q$}

We use the soft cutoff procedure with the dressed RPA boson propagator, Fig.~\ref{fig4}(a), to compute the quantum corrections shown in Fig.~\ref{fig4}(b)-(d). 
In this section, we derive the general form of the corrections, introducing symbolic expressions for the different loop integrals. These integrals depend on the values
of $d_L$ and $d_Q$, through the dimensionality of the loop integral and, more importantly, through the non-perturbative boson propagator, which strongly scales with 
dimension. Combining quantum corrections and re-scaling contributions, we derive general RG equations, which we solve to obtain expressions for critical exponents
of order $1/N_f$ in terms of the loop integrals. These integrals will be evaluated for semi-Dirac fermions ($d_L=d_Q=1$) in Sec.~\ref{sec.exponents_semiDirac} and 
near the upper critical line $2d_L+d_Q=4$ in Sec.~\ref{sec.epsilon}, where we discuss different $\epsilon$ expansions. 

The cutoff independent quantum corrections are obtained by taking the logarithmic derivatives of the diagrams in Fig.~\ref{fig4}, with $ z_L \ell = \log(\Lambda/\Lambda_0)$, 
where the extra factor of $z_L$ comes from the redefinition of ``unit length''. Expanding the fermionic self energy diagram, Fig.~\ref{fig4}(b), to leading order in frequency, momenta 
and $\Delta$, we obtain 
\begin{eqnarray}
\label{eq.fermionself_loops}
\frac{d }{d \ell}\Sigma & = &  -i\, z_L \left[ \delta\Sigma_L \left(k_0\gamma_0+ \v{k}_L\cdot \bm{\gamma}_L\right)\right.\nonumber\\
& & + \left.\left(\delta\Sigma_Q \,v_Q^2 \v{k}_Q^2 +\delta \Sigma_\Delta \,\Delta\right)\gamma_Q \right],
\end{eqnarray}
with certain loop integrals $\delta\Sigma_L, \delta\Sigma_Q, \delta\Sigma_\Delta\sim 1/N_f$ that will be computed later. Likewise, the quantum corrections corresponding to 
the diagrams in Figs.~\ref{fig4}(c) and (d), which renormalize the Yukawa coupling $g$ and order-parameter mass $m^2$, respectively, can be written in the general form 
\begin{equation}
\label{eq.otherloops}
\frac{d}{d \ell} \Xi  =  z_L \delta \Xi \, \frac{g}{\sqrt{N_f}},\quad  \frac{d}{d \ell} \tilde{\Pi}(0) =  z_L \delta\tilde{\Pi} \,m^2.
\end{equation}
Here $\delta \Xi,\delta\tilde{\Pi}\sim 1/N_f$ are one and two-loop integrals over internal momenta. 
Combining these quantum corrections with the rescaling given in Eqs.~(\ref{eq.dLdQresc1}), (\ref{eq.dLdQresc2}) and (\ref{eq.dLdQresc3}), we obtain the following set
of RG equations, 
\begin{eqnarray}
\label{eq.RGvQ}
\frac{d\ln v_Q^2}{d\tilde{\ell}}  & = &  \delta\Sigma_Q -1 - d_L - (2 + d_Q)\frac{z_Q}{z_L}-\Delta_\Psi,\\
\label{eq.RGg}
\frac{d\ln g }{d\tilde{\ell}}& = &  \delta\Xi - 2\left(1+d_L+d_Q\frac{z_Q}{z_L}\right) - \Delta_\Psi - \frac{\Delta_{\phi}}{2},\\
\label{eq.RGDelta}
\frac{d\ln\Delta}{d\tilde{\ell}}& = & \delta \Sigma_\Delta -1-d_L-d_Q\frac{z_Q}{z_L}-\Delta_\Psi = \nu_\Delta^{-1},\\
\label{eq.RGm2}
\frac{d\ln m^2}{d\tilde{\ell}} & = & \delta\tilde{\Pi} - 1 - d_L - d_Q \frac{z_Q}{z_L} -\Delta_{\phi}= \nu_{\phi}^{-1}, 
\end{eqnarray}
where we have defined $\tilde{\ell}=z_L\ell$. The critical dimensions of the fermion and boson fields consist of the tree-level scaling $[\ldots]$, 
given in Eqs.~(\ref{eq.fermion_TL}) and (\ref{eq.boson_TL}),  and
the anomalous dimensions $\eta$, $\Delta_\Psi = [\bar{\Psi}\Psi]+\eta_\Psi$, $\Delta_\phi = [\phi\phi]+\eta_\phi$. Note that the RG flow of the two 
relevant coupling constants $\Delta$ and $m^2$ defines the correlation length exponents $\nu_\Delta$ and $\nu_\phi$ of the multi-critical point. 
In addition to the above RG equations, we have to satisfy the constraint 
\begin{equation}
\label{eq.Deltapsiconstraint}
\Delta_\Psi = \delta\Sigma_L - 2-d_L-d_Q \frac{z_Q}{z_L},
\end{equation}
which follows form the condition that the the coefficient of the linear terms $k_0\gamma_0 + \bm{k}_L\cdot\bm{\gamma}_L$ of the fermion propagator 
remains constant under the RG.  

From the RG equations it is straightforward to extract general expressions for critical exponents in terms of the loop integrals. Inserting Eq.~(\ref{eq.Deltapsiconstraint})
into Eq.~(\ref{eq.RGvQ}) and demanding that $v_Q$ does not flow under the RG, we obtain 
\begin{equation}\label{eqn:d_L-d_Q_zLzQ_ratio}
\frac{z_Q}{z_L} = \frac12 - \frac12\left(\delta\Sigma_L - \delta\Sigma_Q   \right),
\end{equation}
for the ratio of scaling exponents of momenta along quadratic and linear directions.  As to be expected,  $1/N_f$ corrections to the ``tree-level'' value of $1/2$ arise because 
of different fermionic self-energy corrections  along quadratic and linear directions. Using this result and the tree-level scaling dimension of the fermion 
field (\ref{eq.fermion_TL}), we obtain the anomalous dimension of the fermion field, 
\begin{equation}\label{eqn:d_L-d_Q_fermionic_anomalous_dimension}
\eta_\Psi =  \delta \Sigma_L+\frac{d_Q}{2}\left(\delta\Sigma_L-\delta\Sigma_Q\right).
\end{equation}

In order to determine the critical dimension $\Delta_\phi$ of the boson field and the related anomalous dimension $\eta_\phi$, we can use the same argument as for the 
large-$N_f$ GNY theory: since it is possible, to scale out the Yukawa coupling $g$ by the simple transformation $\phi\to\phi/g$ and $m^2\to g^2 m^2$, the coupling 
$g$ should not renormalize. From Eq.~(\ref{eq.RGg}) and the already determined critical exponents we obtain the anomalous dimension
\begin{equation}
\eta_\phi = 2-d_L-\frac{d_Q}{2}+2\left( \delta \Xi - \delta \Sigma_L \right)+ d_Q\left(\delta\Sigma_L-\delta\Sigma_Q\right)
\end{equation}
of the boson fields. Note that $\eta_\phi$ has a contribution of order $(1/N_f)^0$ that vanishes along the upper critical dimension line $2d_L+d_Q=4$. 
In the following we redefine the order parameter scaling such that the anomalous dimension is solely composed of quantum corrections, and the  $(1/N_f)^0$ contribution is absorbed into the tree-level scaling,
\begin{align}
[\phi \phi] &=  -(2 d_L +d_Q),\\
\label{eqn:d_L-d_Q_bosonic_anomalous_dimension}
\eta_\phi &= 2\left( \delta \Xi - \delta \Sigma_L \right)+ d_Q\left(\delta\Sigma_L-\delta\Sigma_Q\right).
\end{align}

And finally, from Eqs. (\ref{eq.RGDelta}) and (\ref{eq.RGm2}), we extract the two correlation length exponents,
\begin{eqnarray}
\nu_\Delta^{-1} & = & 1 +\delta\Sigma_\Delta- \delta\Sigma_L,\\
\nu_{\phi}^{-1} & = & -1+d_L+\frac{d_Q}{2}+ \delta\tilde{\Pi}-2(\delta\Xi -\delta\Sigma_L)\nonumber\\
& & -\frac{d_Q}{2}\left(\delta\Sigma_L-\delta\Sigma_Q\right),
\end{eqnarray}
of the multi-critical fixed point, where $\nu_{\phi}$, $\nu_\Delta$ correspond to the symmetry breaking transition and the topological phase transition, respectively.

\subsection{Exact 1/$N_f$ exponents for semi-Dirac fermions ($d_L=1,d_Q=1$)} 
\label{sec.exponents_semiDirac}

In order to calculate the quantum corrections  $\delta\Sigma_L$, $\delta\Sigma_Q$, $\delta\Sigma_\Delta$, $\delta\Xi$, and $\delta\tilde{\Pi}$ for semi-Dirac fermions, 
we first need to compute the dressed IR boson propagator $G_\phi^{-1} (k) = \Pi(k) +m^2$. Unlike for relativistic fermions, it is not possible to analytically evaluate the 
fermionic polarization $\Pi(k)$ for aniosotropic nodal fermions \cite{nagaosa_semi_dirac_nature,nagaosa_semi_dirac_prl,sur_roy_semi_dirac}.
As shown in Appendix~\ref{app.dLdQA1}, the bosonic self energy for $d_L=d_Q=1$ can be written in the form  
\begin{equation}\label{eqn:SD-RPA_for_SD}
\Pi(k) = \frac{g^2}{8\pi^2} \abs{k_Q}F\left(\frac{k_0^2 + k_L^2}{v_Q^4 k_Q^4}\right),
\end{equation}
where the function $F$ is defined as the integral
\begin{equation}
\label{eq.Funumerics}
F(u) =\int_0^1 dt \int_{-\infty}^{\infty}dp \frac{(p+1)^4-p^2(p+1)^2+(1-t)u}{(p+1)^{4}t +p^4(1-t)+t(1-t)u}.
\end{equation}

Notice that in this form, $\Pi(k)$ still satisfies the condition (\ref{eq.cutoffinddLdQ}) for cutoff independence, $n_\phi=d_L+d_Q/2-1=1/2$, since $\abs{k_Q}\sim\varepsilon^{1/2}$ 
while the argument of the function $F$ is independent of $\varepsilon$.

\begin{figure}[t!]
 \includegraphics[width=\columnwidth ]{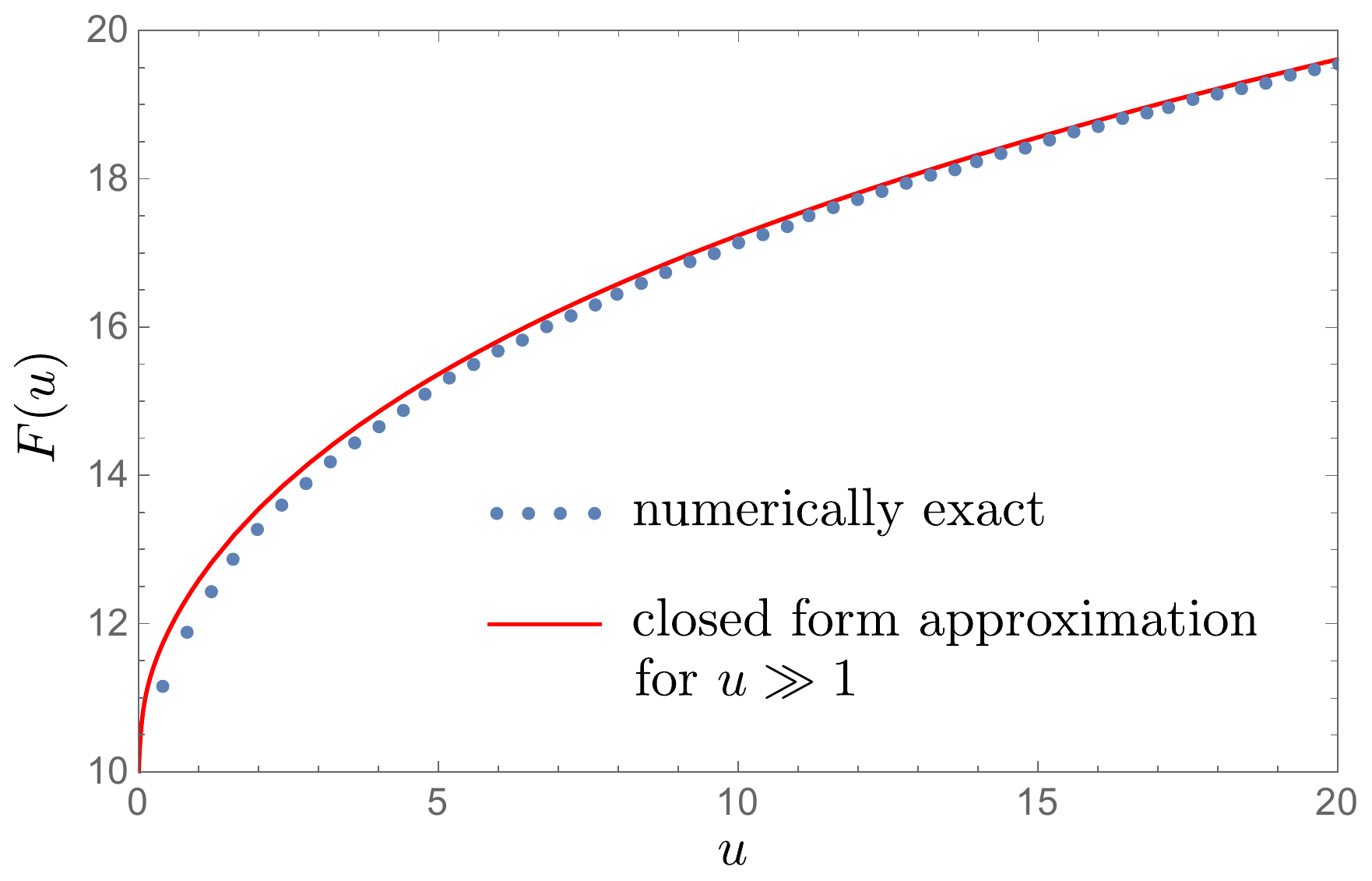}
 \caption{The function $F(u)$ determining the bosonic self energy (\ref{eqn:SD-RPA_for_SD}). The blue dots show the exact result from numerical integration of
 Eq.~(\ref{eq.Funumerics}), the red solid 
 line the closed expression obtained in the regime $u\gg1$.}
\label{fig6}
\end{figure}

The dominant contributions to the quantum corrections come from the regime where $k_Q\to 0$ for finite $k_0$, $k_L$, corresponding to large 
values of the argument $u$. In this regime, it is possible to obtain a closed asymptotic form for $F(u)$, resulting in the approximate boson self energy 
\begin{equation}
\Pi(q)\approx  g^2\left[\frac{a_L(q_0^2 + q_L^2)}{(q_0^2 +q_L^2 + b_Q^4 q^4_Q)^{\frac{3}{4}}  }+ \frac{a_Q q_Q^2 }{(q_0^2 +q_L^2 +b_Q^4 q^4_Q)^{\frac{1}{4}}}\right],
\label{eqn:apxLoopC-asymptotic_rpa}
\end{equation}
with $a_L = \Gamma(5/4)^2 / \sqrt{2} \pi^{3/2}$, $a_Q = 5\Gamma(3/4)^2 / 16\sqrt{2} \pi^{3/2}$, and $b_Q = 8 a_Q$. 

The function $F(u)$ obtained from numerically evaluating the integral (\ref{eq.Funumerics}) and the closed asymptotic approximation for large $u$,  
leading to Eq.~(\ref{eqn:apxLoopC-asymptotic_rpa}),  are shown in Fig.~\ref{fig6}.

As shown in Appendix~\ref{app.dLdQA2}, all quantum corrections can be written as one-dimensional integrals over the function $F(u)$, e.g. 
\begin{equation}
\delta\Sigma_L = \frac{1}{N_f}\int_{0}^{\infty} \frac{du}{(1+u)^2F(u)}=\frac{0.0797}{N_f},
\end{equation}
where we have used the exact form of $F(u)$ to obtain the numerical value. The other quantum corrections are $\delta\Sigma_Q = 0.0214/N_f$, 
$\delta \Sigma_\Delta =0.2755/N_f$, $\delta\Xi = -0.4350/N_f$, $\delta\tilde{\Pi} = -1.0541/N_f$. The resulting exact critical exponents, describing the multi-critical fixed point 
of semi-Dirac fermions, are 
\begin{eqnarray}
\frac{z_Q}{z_L} & = & \frac{1}{2} - \frac{0.0292}{N_f},\\
\eta_\Psi & = & \frac{0.1089}{N_f},\quad \eta_{\phi}  =   -\frac{0.9712}{N_f},\\
\nu_{\phi}^{-1} & = & \frac{1}{2}  -\frac{0.0537}{N_f},\quad \nu_\Delta^{-1} = 1 + \frac{0.1958}{N_f}.
\end{eqnarray}

Numerical values of the quantum corrections obtained with the approximate closed form of $F(u)$, corresponding to the approximate propagator (\ref{eqn:apxLoopC-asymptotic_rpa}),
are given in Appendix~\ref{app.dLdQA2}. These values deviate by less than 3.5\% from the exact ones, except for $\delta\Sigma_Q$ where the deviation is about 17\%.
The larger deviation for $\delta\Sigma_Q$ is due to the fact that the corresponding integral has considerably more weight for small $u$.

\subsection{$\epsilon_L$ and $\epsilon_Q$ expansion below the upper critical line}
\label{sec.epsilon}

We now use the soft cutoff approach to investigate $\epsilon$ expansions below the upper critical line $2d_L+d_Q = 4$ of anisotropic $d_L$-$d_Q$ nodal fermion systems.
In principle, such expansions should allow for a controlled descent to strongly interacting systems of interest, such as the semi-Dirac fermion system with $d_L=d_Q=1$.  

There is however a freedom in the choice of the starting point on the upper critical line. Here we focus on two natural starting points which correspond to a descent towards 
semi-Dirac fermions by expanding in the number of (i) linear and (ii) quadratic dimensions, as illustrated in Fig.~\ref{fig5}. This corresponds to (i) $d_L=(3-\epsilon_L)/2,\ d_Q=1$
and (ii) $d_L=1,\ d_Q=2-\epsilon_Q$, where semi-Dirac fermions are reached for $\epsilon_L=1$ and $\epsilon_Q=1$, respectively. 
Note that the expansion (ii) in the number of quadratically dispersing directions was used in Ref.~\cite{sur_roy_semi_dirac}.

As explained in Sec.~\ref{subsec:dL_dQ_RPA}, the bare order parameter propagator, $G_{\phi,0}^{-1}(q)\sim q^2$, does not satisfy the condition for cutoff independence, even at the 
upper critical line. A fully perturbative RG calculation as the one used in the $D=4-\epsilon$ expansion for GNY theory is therefore insufficient.
Instead, the phenomenon of Landau damping remains of crucial importance at the upper critical line, giving rise to a non-analytic bosonic self-energy 
 correction $\Pi(q_0,\bm{q}_L,\bm{q}_Q=0)\sim (q_0^2+\bm{q}_L^2)^{1/2}$ along the linear directions. As expected from the condition of cutoff independence, 
 the scaling of the self-energy along the quadratic directions approaches the form $\Pi(q_0=0,\bm{q}_L=0,\bm{q}_Q)\sim \bm{q}_Q^2$.
 
 However, the coefficient diverges logarithmically on approach of the upper critical line. As shown in Appendix~\ref{app:d_L-d_Q-RPA_derivation}, it is possible to extract the 
 leading $1/\epsilon$ pole associated with this divergence. Since the fermionic polarization diagram cannot be calculated for general $q=(q_0, \bm{q}_L, \bm{q}_Q)$
 we approximate the dressed IR boson propagator by the sum of the two asymptotic forms of the self energy along linear and quadratic directions.

For the expansion (i) in the linear dimensions, $d_L=(3-\epsilon_L)/2,\ d_Q=1$, we obtain
\begin{equation}\label{eqn:dLdQEps-RPA_epsilonL}
G_{\phi}^{-1}(q) = \frac{g^2\pi^{1/4}}{32 \Gamma(3/4)}(q_0^2+ \bm{q}_L^2)^{1/2} + \frac{1}{\epsilon_L}\frac{g^2}{2\pi^{5/4}\Gamma(1/4)}\bm{q}_Q^2,
\end{equation}
while for the expansion (ii) in the quadratic directions,  $d_L=1,\ d_Q=2-\epsilon_Q$, the result is 
\begin{equation}\label{eqn:dLdQEps-RPA_epsilonQ}
G_{\phi}^{-1}(q) = \frac{g^2}{64}(q_0^2+ \bm{q}_L^2)^{1/2} + \frac{1}{\epsilon_Q}\frac{g^2}{8\pi^{2}}\bm{q}_Q^2.
\end{equation}
Details of the derivation can be found in Appendix~\ref{app:d_L-d_Q-RPA_derivation}. Note that the propagators satisfy the condition (\ref{eq.cutoffinddLdQ}) of 
cutoff independence.

\begin{table}
\caption{Critical exponents $z_Q/z_L$, $\eta_\Psi$ and $\eta_\phi$ for two distinct $\epsilon$-expansions around the upper critical line $2d_L+d_Q=4$. 
Here $\alpha_1^{-1} = (2\pi)^{1/4}\Gamma(9/4)$ and $\alpha_2=\pi^2/8$, for brevity.}\label{table:epsilon_exponents}
\begin{tabular}[t]{c  c  c}
\hline \hline
 & $\quad d_L=(3-\epsilon_L)/2,\;d_Q=1$ &   $\quad d_Q=2-\epsilon_Q,\;d_L=1$ \\ \hline
 $\frac{z_Q}{z_L}$       &  $\quad \frac12  + \alpha_1 \frac{\sqrt{\epsilon_L}}{N_f}- \frac{3\epsilon_L}{2 N_f} $ & $\quad \frac12-\frac{\epsilon_Q}{2N_f} \log(\alpha_2 \epsilon_Q)-\frac{5\epsilon_Q}{4N_f}  $  \\ 
$\eta_\Psi$ & $\quad-\alpha_1 \frac{\sqrt{\epsilon_L}}{2N_f}+ \frac{3\epsilon_L}{2N_f}$ & $\quad \frac{\epsilon_Q}{N_f}\log(\alpha_2\epsilon_Q)+ \frac{3\epsilon_Q}{N_f}$ \\ 
$\eta_{\phi}$ & $\quad  - \alpha_1 \frac{8\sqrt{\epsilon_L}}{N_f}+\frac{5\epsilon_L}{N_f} $ & $\quad \frac{4\epsilon_Q}{N_f} \log(\alpha_2\epsilon_Q)+\frac{4\epsilon_Q}{N_f} $ \\ 
\hline
\end{tabular}
\end{table}

To demonstrate the fundamental differences between the $\epsilon_L$ and $\epsilon_Q$ expansions, it is sufficient to calculate the ratio of the scaling dimensions $z_Q/z_L$ and the 
anomalous dimensions $\eta_\Psi$ and $\eta_\phi$ for the two cases. These critical exponents are expressed in 
Eqs.~(\ref{eqn:d_L-d_Q_zLzQ_ratio},\ref{eqn:d_L-d_Q_fermionic_anomalous_dimension},\ref{eqn:d_L-d_Q_bosonic_anomalous_dimension}) in terms of the quantum corrections $\delta\Sigma_L$, 
$\delta\Sigma_Q$, which arise from the expansion of the fermion self energy [Fig.~\ref{fig4}(b)], and $\delta\Xi$ from the vertex correction [Fig.~\ref{fig4}(c)].
The corresponding one-loop integrals are computed in Appendix~\ref{app:dL-dQ-epsilon_corrections}, using the the soft cutoff approach with the dressed IR boson 
propagators Eqs. (\ref{eqn:dLdQEps-RPA_epsilonL}) and (\ref{eqn:dLdQEps-RPA_epsilonQ}). The resulting critical exponents are summarized in Table~\ref{table:epsilon_exponents}.

For the expansion along the number of quadratic dimensions, $d_L=1,\ d_Q=2-\epsilon_Q$, we find that the quantum corrections computed with the soft-cutoff RG and the 
non-perturbative boson propagator (\ref{eqn:dLdQEps-RPA_epsilonQ}) are in perfect agreement with those obtained in Ref.~\cite{sur_roy_semi_dirac}, when evaluated at the interacting fixed 
point. To leading order in $\epsilon_Q \log\epsilon_Q$ the critical exponents also agree, once the different definitions of the critical dimension of the bosonic field and the number of fermionic 
flavors have been accounted for. Further details can be found in Appendix~\ref{app.scaling_comp}.

While  the leading quantum corrections are non-analytic for both the $\epsilon_L$ and $\epsilon_Q$ expansions, the functional dependencies $\sim\sqrt{\epsilon_L}$ and 
$\sim\epsilon_Q \log\epsilon_Q$ are completely different, potentially signaling an intrinsic problem with $\epsilon$ expansions in $d_L$-$d_Q$ nodal fermion systems. 
This is further supported by the significant disparity between critical exponents obtained from the extrapolation of the two expansions to the semi-Dirac point, $\epsilon_L=1$
and $\epsilon_Q=1$.

\section{Discussion}\label{sec:discussion}

We have investigated the universal critical behavior of topological nodal point semimetals at quantum phase transitions that are driven by strong local interactions. We have 
developed a soft cutoff RG approach that can be used to calculate exact critical exponents to leading order $1/N_f$ in experimentally relevant spatial dimensions. 

At the heart of the problem is the phenomenon of Landau damping of order parameter fluctuations by gapless fermion excitations. This leads to non-analytic bosonic self-energy 
corrections which dominate over the bare boson propagator in the IR long-wavelength limit. Landau damping is therefore essential for the universal critical behavior of the system. 
The phenomenon of Landau damping is inherently non-perturbative and not captured by  perturbative RG schemes  that are based upon the successive decimation of UV modes
\cite{nagaosa_semi_dirac_prl}. 

As demonstrated within our soft cutoff approach, not accounting for Landau damping, or more generally, using an incorrect IR boson propagator, leads to non-universal results that
depend on the choice of the UV cutoff scheme.  In turn, enforcing that the quantum corrections do not depend on the cutoff function and on which frequency and momentum directions
the cutoff acts upon, the correct IR scaling of the Landau damped boson propagator can be deduced. These scaling constraints are satisfied by the fully dressed RPA boson propagator.

Our soft cutoff approach unifies all possible cutoff schemes, including those based on cylindrical and spherical hard cutoff momentum shells. Our work therefore 
demonstrates  that \textit{any} RG scheme is valid and will produce the same universal results, given that the correct IR boson propagator is used. 
This should resolve controversies over the ``correct'' RG shell schemes when there are quantitative discrepancies in the literature, such as in the case of 
 double-Weyl semimetals \cite{lai_double_weyl_2015,hong_yao_double_weyl_2015}.

Using the soft cutoff RG with the non-perturbative RPA boson propagator, we have computed the exact critical exponents to leading order $1/N_f$ for relativistic Weyl or 
Dirac fermions as well as for two-dimensional anisotropic semi-Dirac fermions, coupled to an Ising order parameter field. The soft cutoff method has a clear advantage 
over hard cutoff schemes,  as it significantly simplifies the calculation of diagrams beyond one-loop order. 
 For the well studied relativistic case, the soft cutoff RG indeed reproduces the exact critical exponents obtained by conformal 
bootstrap \cite{Vasilev_nu_n2_1993,Gracey94,Iliesiu+18} and other 
field-theoretical techniques \cite{Gat1990,Gracey_eta_n2_1991,Gracey_nu_n2_1992}.     

We briefly compare some of our exact critical exponents for semi-Dirac fermions,
\begin{equation}
\eta_\Psi  =  \frac{0.1089}{N_f}, \quad \eta_{\phi}  =   -\frac{0.9712}{N_f}, \quad \nu_{\phi}^{-1} =  \frac{1}{2}  -\frac{0.0537}{N_f},
\notag
\end{equation}
with those reported in the literature. As discussed in Appendix~\ref{app.scaling_comp} one needs to account for different definitions of the 
number of fermion flavors $N_f$ and scaling exponents $z_L$ and $z_Q$. 

Ref.~\cite{uryszek_semi_dirac} employed one-loop perturbative RG with the bosonic IR divergence in $c_L$~\eqref{eq.bareboson} regulated by the RPA re-summation. 
Although a stable interacting critical fixed point was located, the results are inherently cutoff dependent as the partially dressed boson propagator does not satisfy the 
scaling constraint \eqref{eq.cutoffinddLdQ} for cutoff independence. Moreover, the two-loop diagrams that contribute to the mass renormalization and hence the correlation
length exponent $\nu_\phi$ at order $1/N_f$ were neglected.  
Consequently, there are significant discrepancies in the exponents $\eta_\Psi = 0.0229/N_f$, $\eta_\phi = -0.1004/N_f$ and  $\nu^{-1}_{\phi} = 1/2+ 0.2466/N_f$. 

Ref.~\cite{Wang+17} obtained cutoff independent fermion self energy quantum corrections, using an approximation for the bosonic self energy $\Pi(q)$ that satisfied 
the scaling constraint \eqref{eq.cutoffinddLdQ}. However, the approximation did not capture the full anisotropy, resulting in $\delta \Sigma_Q$ that is only $12\%$ of that 
found here (\ref{eq.SigQcomp}). In addition, the renormalization of $z_L/z_Q$ was not accounted for in scaling, resulting in $\eta_\Psi = 0.0870/N_f$. Other exponents 
and quantum corrections were not computed.

We have compared $\epsilon$ expansions that descended on the semi-Dirac point by expanding in the number of (i) linear $d_L=(3-\epsilon_L)/2,\ d_Q=1$ and 
(ii) quadratic $d_L=(3-\epsilon_L)/2,\ d_Q=1$ dimensions. In both cases we found  quantum corrections and critical exponents that are non-analytic in $\epsilon$. 
However, the functional dependencies $\sqrt{\epsilon_L}$ and $\epsilon_Q \log \epsilon_Q$ are completely different. 
This calls into question the validity, uniqueness, and extent of perturbative control of this approach for anisotropic nodal fermion systems, in contrast to relativistic GNY 
and bosonic $\phi^4$ theories. Further analysis, and exploration at higher loop order is required to make concrete conclusions. 
Finally, it would be interesting to study the crossover behavior from perturbative $\epsilon$ to integer $d_L$-$d_Q$ systems, similar to recent work on quantum critical metals~\cite{schlief_epsilon_large_N_afm_PRB}. There it was found that low energy and integer dimension limits do not commute.

The scaling constraints we have derived from the requirement of cutoff independence highlight the important role of non-perturbative effects in quantum critical systems.
Interestingly, it has been known for some time that the non-perturbative screening of long-range Coulomb interactions in relativistic nodal systems is crucial \cite{vozmediano_honeycomb_1994,vozmediano_graphene_columb1999,dtSon_coulomb_graphene}, and formally equivalent to Landau damping.
However, such effects are typically neglected when studying spontaneous symmetry breaking from short range interactions.
As a result, cutoff independence is often violated in the literature when studying the quantum criticality of two dimensional 
systems \cite{uryszek_semi_dirac,hong_yao_fermion_induced_QCP_2017_NatCom,hong_yao_fermion_induced_QCP_Dirac_2D_PRB_2017,christou_Dirac_gauge_fields_2020,bruno_nodal_line}.
Screening is also important in anisotropic semi-Dirac systems at low energies. Our analysis  shows that the entire polarization function is relevant, leading us to agree with Ref.~\cite{nagaosa_semi_dirac_prl} regarding Coulomb 
quantum criticality of semi-Dirac fermions: the dynamical part of the polarization should not be neglected, contrary to what was argued in Ref.~\cite{Cho+16}.

There are a number of interesting avenues for future research into strongly interacting nodal systems away from their upper critical dimension. Closely linked to semi Dirac 
fermions are two and three dimensional nodal line semimetals. These are are described by the same effective  field theory (\ref{eq.L_dL_dQ}) but at finite $\Delta<0$, away 
from the topological phase transition point. The criticality of such systems due to spontaneous symmetry breaking was previously studied within 
perturbative RG \cite{bruno_nodal_line}, not taking into account the effects of Landau damping. In nodal line semimetals Landau damping is expected to have even stronger effects 
than in the nodal point case, due to the greatly enhanced electronic density of states at low energies. 
It would also be interesting to revisit nematic quantum phase transitions in quadratic band touching systems, previously studied
within perturbative RG near the upper critical dimension \cite{janssen_herbut_qbt_nematic,herbut_janssen_qbt_top_mott_QBT}.  
The bare tensorial order parameter propagator of these theories does not satisfy the condition of cutoff independence in physically relevant dimensions, highlighting that  
non-perturbative effects are crucial for the universal critical behavior. 

Starting with relativistic dynamics, the presence of emergent gauge fields coupled to order parameter fields 
\cite{huh_sachdev_nematic_d_wave_SC, christou_Dirac_gauge_fields_2020} can 
alter the dynamical scaling at quantum critical points. In this case, the divergencies associated with non-invertible, damped gauge field propagators can be repaired 
with non-analytic gauge fixing, as is implemented in pseduo-QED \cite{Marino+1993}.
There are instances where broken symmetry states on lattices allow for cubic terms in the order parameter fields in the low energy effective field theory \cite{hong_yao_fermion_induced_QCP_2017_NatCom,hong_yao_fermion_induced_QCP_Dirac_2D_PRB_2017,schererherbutprb2016,christou_Dirac_gauge_fields_2020}. In principal, following the Landau criterion, these can render quantum 
phase transitions first order. However gapless fermion excitations are expected to render such cubic terms irrelevant. In these problems, damping effects away from upper critical 
dimensions have not been accounted for. 
Finally, the interplay of Landau damping and disorder in topological nodal semimetals is largely unexplored. Interestingly,  even 
the stability of the critical fixed point in the GNY theory against weak disorder remains controversial \cite{Nandkishore+13,Yerzhakov+18}.   

In topological nodal systems the fermions are fundamental in the region of the nodal points. 
The fermion dynamics must therefore be analytic, implying that non-analytic fermion self energy corrections are absent. 
Consequently the large $N_f$ expansion is controlled, as $N_f$ does not appear in the fermion propagator. 
This is in contrast to the case of metallic quantum critical systems, in which the fermions are strongly renormalized by the infinite sea of excitations, resulting in non-analytic fermion self energy corrections. 
These terms are typically more relevant than the bare fermion dynamics, and render the large $N_f$ uncontrolled~\cite{ssLee_fermi_surf_2+1}. 
Entirely non-perturbative solutions of the Schwinger-Dyson equations are then required~\cite{Schlief+2017}. 
Using our soft cutoff formalism, it might be possible to derive scaling constraints for both bosonic and fermionic self energies. This could potentially provide an important 
step towards the discovery of such non-perturbative solutions.

\acknowledgements

We thank J\"org Schmalian and Pavel Volkov for fruitful discussions. E.~C. and F.~K. acknowledge financial support from EPSRC under Grant EP/P013449/1.

\appendix

\section{cutoff scheme independent integrals}

\subsection{Soft cutoff integrals at external $q$}
\label{app.soft1}

It can be shown in general that any cutoff independent quantum correction with external momenta from cutoff functions vanishes. This justifies 
why the external momenta appearing in cutoff functions can be set to zero. Here we explicitly show this for the only relevant case of
the self-energy correction 
\begin{equation}
\frac{d}{d\ell} \Sigma(q) =  -\Lambda \frac{d}{d\Lambda} \frac{g^2}{N_f} \int_k G_\Psi (k+q)   A_{k+q} G_\phi(k)A_k,
\end{equation}
which renormalizes the fermion propagator. In principle, terms with linear $q$ dependence could arise from the external 
momentum in either $G_\Psi (k+q)$ or $A_{k+q}$. To show that the latter contributions vanish, we set $G_\Psi (k+q)=G_\Psi (k)$
and expand $A_{k+q}=A\left(a_\mu  (k_\mu+q_\mu)^2/\Lambda^2 \right)$ to linear order in $q_\mu$,
\begin{equation}
A_{k+q} -A_k = 2  A'\left(\frac{f y^2}{\Lambda^2}   \right)  \frac{y}{\Lambda^2} a_\nu \hat{k}_\nu q_\nu.
\end{equation}
Here we have substituted $k=y\hat{k}$, $\hat{k}_\mu \hat{k}_\mu =1$, and introduced the angular function $f=f(\hat{\Omega})=a_\mu  \hat{k}_\mu^2$.
Using that $G_\Psi (k) = i \hat{k}_\mu\gamma_\mu/y$ and $G_\phi(k) = G_\phi(\hat{k})/y^{n_\phi}$, where $n_\phi = D-2$ to ensure cutoff independence, the 
linear $q$ correction arising from $A_{k+q}$ is equal to 
\begin{equation}
\begin{gathered}
-\Lambda \frac{d}{d\Lambda} \frac{g^2}{N_f} \int_k G_\Psi (k)   A_{k+q} G_\phi(k)A_k \\
= -i \frac{g^2}{N_f} q_\mu \gamma_\mu \int_{\hat{\Omega}} a_\mu \hat{k}_\mu^2 G_\phi(\hat{k}) I(\hat{\Omega}),
\end{gathered}
\end{equation}
with a radial integral 
\begin{equation}
I(\hat{\Omega}) =  \Lambda \frac{d}{d\Lambda} \int_0^\infty dy \frac{2y}{\Lambda^2}A'\left(\frac{f(\hat{\Omega}) y^2}{\Lambda^2}   \right) 
A\left(\frac{f(\hat{\Omega}) y^2}{\Lambda^2}   \right).
\end{equation}
It is straightforward to show that $I(\hat{\Omega}) = 0$,
\begin{equation}
I(\hat{\Omega}) = \frac{1}{2f(\hat{\Omega})} \Lambda \frac{d}{d\Lambda} \left[ A^2\left(\frac{f(\hat{\Omega}) y^2}{\Lambda^2}   \right)   \right]_{y=0}^{y=\infty} = 0.
\end{equation}
The last step simply follows from the boundary conditions of the cutoff function, $\lim_{z\to\infty} A(z) =0$ and $A(0)=1$.

\subsection{Proof of Eq.~(\ref{identity})}
\label{app.soft2}

\noindent
We define the function
\begin{equation}
F\left(\frac{y^2}{\Lambda^2}\right) := \prod_i A^{n_i}\left(\frac{y^2 f_i}{\Lambda^2}\right).
\end{equation}

From the boundary conditions of the cutoff function $A(z)$, Eq.~(\ref{eq.Alimits}), it follows 
that $F(0)=1$ and $\lim_{z\to\infty} F(z) = 0$. Evaluating the left-hand side of Eq.~(\ref{identity}),
we obtain
\begin{equation}
\begin{gathered}
\Lambda \frac{d}{d\Lambda} \int_0^\infty \frac{dy}{y} F\left(\frac{y^2}{\Lambda^2}\right)  =  - \int_0^\infty \frac{dy}{y} F'\left(\frac{y^2}{\Lambda^2}\right)\frac{2y^2}{\Lambda^2}\nonumber\\
= -  \int_0^\infty dy \,\frac{d}{dy}\left[ F\left(\frac{y^2}{\Lambda^2}\right)  \right] = -\left[ F\left(\frac{y^2}{\Lambda^2}\right)  \right]_{y=0}^{y=\infty} = 1,
\end{gathered}
\end{equation}
which proves Eq.~(\ref{identity}). If on the other hand, the integrand scales as $1/y^\alpha$ with $\alpha\neq 1$, the logarithmic derivative of the integral won't be independent of 
the cutoff scale $\Lambda$ and the choice of cutoff function $A$.

\section{Useful integral identities}

The loop calculations utilize integral identities that follow from the integral representations of the $\Gamma$ function.
Typically, $k$-integrals are rewritten in hyper-spherical coordinates $k_\mu = y \hat{k}_\mu$ with $\hat{k}_\mu\hat{k}_\mu=1$,
\begin{equation}
\int_k = \int \frac{d^D k}{(2\pi)^D} = \int \frac{d\hat{\Omega_k}}{(2\pi)^D} \int_0^\infty dy \,y^{D-1}.
\end{equation}
The radial integral identity
\begin{equation}
\int_{0}^{\infty} dy \frac{y^{D-1+\alpha}}{\left(y^\beta + M\right)^n} = \frac{\Gamma(\frac{D+\alpha}{\beta} )\Gamma(n-\frac{D+\alpha}{\beta} )}{\beta\,\Gamma( n) M^{n-\frac{D+\alpha}{\beta}}},
\label{eqn:apxLoopC-I_radial}
\end{equation}
is valid for $D+\alpha>0$ and $n \beta > D+ \alpha$.
The angular integral identity over the $D$-dimensional unit sphere,
\begin{equation}
\int \frac{d\hat{\Omega}_k}{(2\pi)^D} \hat{k}_\mu^{2n} = S_D \frac{\Gamma(\frac{D}{2})\Gamma(\frac{2n+1}{2})}{\sqrt{\pi} \Gamma (\frac{2n+D}{2})},
\label{eqn:apxLoopC-I_angular}
\end{equation}
for integer $n$. Integrals over odd powers of $\hat{k}_\mu$ are zero, by symmetry. Here $\mu=0,\dots,D$ is not summed over and
\begin{equation}
\label{eq.volD}
S_D = \frac{1}{(2\pi)^\frac{D}{2}} \frac{2\pi^\frac{D}{2}}{\Gamma(\frac{D}{2})}
\end{equation}
is the surface area of a $D$-dimensional unit sphere.

\medskip
\noindent
The Feynman parameterization,
\begin{equation}
\frac{1}{a^n b^m}  = \frac{\Gamma(n+m)}{\Gamma(n)\Gamma(m)} \int_0^1 dt \frac{t^{n-1}(1-t)^{m-1}}{[t a+(1-t)b]^{n+m}},
\label{eqn:apxLoopC-I_feynman}
\end{equation}
is conjunction with appropriate linear momentum shifts is used to render integrals radially symmetric. 

\medskip
\noindent
The Feynman parameter integral identity,
\begin{equation}
\int_0^1 dt\ t^{a} (1-t)^{b} = \frac{\Gamma\left(a+1\right)\Gamma\left(b+1\right)}{\Gamma\left(a+b+2\right)},
\label{eqn:apxLoopC-I_param}
\end{equation} 
is valid for $a>-1$ and $b>-1$.

\section{GNY diagrams}

\subsection{RPA boson propagator}
\label{app.GNY1}

\noindent
We work in units where the Fermi velocity $v=1$, such that 
\begin{equation}
\label{eq.Fermiprop}
G_\Psi(k)= \frac{i k_\mu \gamma_\mu}{k^2}.
\end{equation}
The fermion loop diagram is displayed in Fig.~\ref{fig3}(a).
We calculate the regularized fermion loop $\Pi(q)\rightarrow \Pi(q) - \Pi(0)$,
\begin{align}
\Pi(q) &= \frac{g^2}{N_f} \int_k \text{tr}\,\left[ G_\Psi(k+q) G_\Psi(k)-G_\Psi(k) G_\Psi(k)\right]\notag\\
&=g^2 \int_k  \frac{(k_\mu+q_\mu)q_\mu}{(k+q)^2 k^2},
\end{align}
where we have used that $\text{tr}\,\gamma_\mu \gamma_\nu = N_f \delta_{\mu\nu}$. After 
using the Feynman parametrization (\ref{eqn:apxLoopC-I_feynman}) with $n=m=1$, $a=(k+q)^2$ 
and $b=k^2$, and substituting $\tilde{k} = k+tq$, the $\tilde{k}$ integral is radially symmetric, 
\begin{equation}
\Pi(q) = g^2\int_0^1 dt  \int_{\tilde{k}} \frac{(1-t)q^2}{[\tilde{k}^2 + t(1-t) q^2)]^2}.
\end{equation}

Evaluating the $\tilde{k}$ integral, using the radial integration formula (\ref{eqn:apxLoopC-I_radial}), and then 
carrying out the one-dimensional integral over the Feynman parameter $t$, using the identity (\ref{eqn:apxLoopC-I_param}), 
we obtain
\begin{equation}
\Pi(q) = \frac{g^2S_D \alpha_D }{v^{D-1}} (q_0^2 + v^2 \v{q}^2)^{\frac{D-2}{2}},
\end{equation}
where we have reinstated the Fermi velocity $v$, and defined
\begin{equation}
\alpha_D=   -\frac{\pi}{2\sin(\frac{\pi D}{2})} \frac{ \Gamma(D/2)^2} { \Gamma(D-1)}.
\end{equation}

\noindent
The resulting dressed RPA boson propagator is given by 
\begin{eqnarray}
G_\phi^{-1}(q) & = &  G_{\phi,0}^{-1}(q) + \Pi(q) \nonumber\\
& = &   \frac{g^2S_D \alpha_D }{v^{D-1}} (q_0^2 + v^2 \v{q}^2)^{\frac{D-2}{2}} +m^2,
\end{eqnarray}
where we neglected the sub-leading momentum and frequency terms from the bare propagator $G_{\phi,0}^{-1}(q)$.

\subsection{Soft cutoff one-loop quantum corrections}
\label{app.GNY2}

In the following we work in rescaled units, such that $v=1$. The dependence on the Fermi velocity will be reinstated in the end. 
The propagators are augmented by cutoff functions $A$ as described in the main text with $A_k = A(a_\mu k_\mu^2/\Lambda^2)$.
The flow of the fermion self energy correction $\Sigma(q)$, Fig.~\ref{fig4}(b), is
\begin{equation}
\frac{d}{d\ell} \Sigma(q) =-\Lambda \frac{d}{d\Lambda} \frac{g^2}{N_f} \int_k  G_\Psi (k+q) G_\phi(k) A_{k}A_k.
\end{equation}
We extract the relevant linear $q$ term on the critical surface $m^2=0$,
\begin{equation}
\frac{d}{d\ell} \Sigma(q) = \frac{i q_\mu \gamma_\nu}{S_D \alpha_D N_f} \Lambda \frac{d}{d\Lambda}\int_k  \frac{ 2 k_\mu k_\nu-\delta_{\mu \nu}k^2}{k^{D+2}}A_k^2,
\end{equation}
and rewrite the integral in terms of angular and radial integrals, defining $k=y\hat{k}$,
\begin{eqnarray}
\frac{d}{d\ell} \Sigma(q) & = &  \frac{i q_\mu \gamma_\nu}{S_D \alpha_D N_f} \int_{\hat{\Omega}} \left( 2 \hat{k}_\mu \hat{k}_\nu - \delta_{\mu \nu} \right)\nonumber\\
& & \times \Lambda \frac{d}{d\Lambda} \int_0^\infty \frac{dy}{y} A^2\left(\frac{f(\hat{\Omega})y^2}{\Lambda^2} \right).
\end{eqnarray}
While the radial $y$ integral becomes trivial, using the soft cutoff integral identity (\ref{identity}), the angular integral can be computed using Eq.~(\ref{eqn:apxLoopC-I_angular}).
The final result is 
\begin{equation}
\frac{d}{d\ell} \Sigma(q)  =   -i\frac{D-2}{\alpha_D D N_f}(q_0 \gamma_0 + v \v{q} \cdot \v{\gamma}).
\end{equation}

To compute the flow of the vertex correction $\Xi$, Fig.~\ref{fig4}(c), we follow the same steps, 
\begin{eqnarray}
\frac{d}{d\ell} \Xi & = &  \Lambda \frac{d}{d\Lambda} \frac{g^3}{\sqrt{N_f}^3} \int_k     G^2_\Psi(k) G_\phi(k) A_k^3 \nonumber\\
& = & -\frac{g}{S_D \alpha_D\sqrt{N_f}^3} \Lambda \frac{d}{d\Lambda}\int_{\hat{\Omega}} \int_0^\infty \frac{dy}{y} A^3\left(\frac{f(\hat{\Omega})y^2}{\Lambda^2} \right)\nonumber\\
& = & -\frac{1}{\alpha_DN_f} \frac{g}{\sqrt{N_f}}.
\end{eqnarray}

\subsection{Soft cutoff two-loop quantum corrections}
\label{app.GNY3}

The flow of the two loop boson self energy $\tilde{\Pi}$, Fig.~\ref{fig4}(d), that renormalizes the boson mass (at zero external momentum) is
\begin{align}
\frac{d}{d\ell} &\tilde{\Pi} =  \Lambda \frac{d}{d\Lambda}\frac{g^4}{N_f^2} \int_{k,q} G_\phi(q) A_q \times \text{tr}\, \Big[  \notag\\
&\phantom{=}\,G_\Psi(k+q) G_\Psi(k+q) G_\Psi(k)G_\Psi(k)A^2_{k+q} A^2_k\notag\\
&\phantom{=}\,+2G_\Psi(k+q) G_\Psi(k) G_\Psi(k)G_\Psi(k)A_{k+q} A^3_k\Big].
\end{align}
We extract the relevant $m^2$ contribution
\begin{align}
\frac{d}{d\ell} \tilde{\Pi} &= -\frac{m^2}{(S_D\alpha_D)^2 N_f}  \Lambda \frac{d}{d\Lambda}  \int_{k,q} \frac{1}{q^{2D-4} k^2 (k+q)^2 } \notag\\
&\times \Big[ A_k^2 A_{k+q}^2 A_q + \frac{2 (k_\mu+q_\mu)k_\mu}{k^2} A_k^3 A_{k+q} A_q \Big].
\end{align}
The two loop calculation involves more steps.
We use the transformation $q_\mu = y \hat{q}_\mu$, $k_\mu = y x \hat{k}_\mu$, where $\hat{q}_\mu\hat{q}_\mu=1$ and $\hat{k}_\mu\hat{k}_\mu=1$, such that
\begin{align}
\frac{d}{d\ell} \tilde{\Pi} &= -\frac{m^2}{(S_D\alpha_D)^2 N_f}    \int \frac{\hat{\Omega}_k}{(2\pi)^D}\int\frac{d\hat{\Omega}_q}{(2\pi)^D} \int_0^\infty dx\, x^{D-1} \notag\\
&\times \Lambda \frac{d}{d\Lambda}\int_0^\infty  \frac{dy}{y} \frac{1}{x^2(x \hat{k}+\hat{q})^2} \Big[ A_{y x \hat{k}}^2 A_{y (x \hat{k}+\hat{q})}^2 A_{y\hat{q}}\notag\\
&\phantom{=} + \frac{2 (x^2 + x \hat{k}_\mu\hat{q}_\mu)}{x^2} A_{y x \hat{k}}^3 A_{y (x \hat{k}+\hat{q})} A_{y\hat{q}} \Big].
\end{align}
The $y$ integral is evaluated with the soft cutoff identity (\ref{identity}). 
The integral is rendered radially symmetric in $x$ after the introduction of the Feynman parameter $t$ (\ref{eqn:apxLoopC-I_feynman}), with the shift 
$\hat{k} \rightarrow \hat{k}-  t\hat{q}/x$. Then the angular integrals are evaluated, resulting in
\begin{align}
\frac{d}{d\ell} \tilde{\Pi} &= -\frac{m^2}{\alpha_D^2 N_f}  \int_0^1 dt \int_0^\infty  dx \,x^{D-1}\Big\{\notag \\
& \frac{1}{[x^2+t(1-t)]^2}  +4 (1-t) \frac{ x^2 - t(1-t)}{[x^2+t(1-t)]^3}\Big\}.
\end{align}
The radial $x$ integral and the integral over the Feynman parameter $t$ are evaluated using Eqs. (\ref{eqn:apxLoopC-I_radial}) 
and (\ref{eqn:apxLoopC-I_param}), respectively. 
The final result is
\begin{equation}
\frac{d}{d\ell} \tilde{\Pi} = \frac{D-1}{\alpha_D^2 N_f \sin(\frac{\pi D}{2})} \frac{\pi \Gamma(\frac{D}{2})^2}{\Gamma(D-1)}m^2.
\end{equation}

\section{Diagrams for semi-Dirac ($d_L=d_Q=1$) systems}

\noindent
We work in units where  $v_Q=1$, and define
\begin{equation}
\varepsilon_\mu(k) = (k_0,k_L,k_Q^2),
\end{equation}
as well as $\gamma_\mu =( \gamma_0,\gamma_L,\gamma_Q)$, such that (at $\Delta=0$)
\begin{equation}
G_\Psi(k) =i\frac{\varepsilon_\mu(k) \gamma_\mu}{\varepsilon^2(k)}=i \frac{ k_0 \gamma_0 + k_L\gamma_L + k_Q^2 \gamma_Q}{k_0^2 + k_L^2 + k_Q^4}.
\end{equation}

\subsection{RPA boson propagator}
\label{app.dLdQA1}

We compute the boson self energy, which is given by the fermion polarization diagram, 
\begin{eqnarray}
\label{eq.regboson}
\Pi(q) & = & \frac{g^2}{N_f} \int_k \text{tr}\,\left[ G_\Psi(k+q) G_\Psi(k)-G_\Psi(k) G_\Psi(k)\right]\nonumber\\
& = &  g^2 \int_k \Big\{q_0(k_0+q_0)+q_L(k_L+q_L)+(k_Q+q_Q)^2\nonumber\\
& & \times\left[(k_Q+q_Q)^2-k_Q^2\right]\Big\}\Big/\Big\{ (k_0^2+k_L^2+k_Q^4)\nonumber\\
& & \times\left[(k_0+q_0)^2+(k_L+q_L)^2+(k_Q+q_Q)^4\right]\Big\},
\end{eqnarray}
where we have used that $\textrm{tr}\gamma_\mu\gamma_\nu = N_f \delta_{\mu\nu}$. This integral can be rendered radially symmetric
in $(k_0,k_L)$ by introducing the Feynman parameter $t$ (\ref{eqn:apxLoopC-I_feynman}) followed by the shift $(k_0,k_L) \rightarrow (k_0,k_L) - t (q_0,q_L)$, 
\begin{align}
& \Pi(q)  =  \frac{g^2}{4\pi^2} \int_0^1 dt \int_{-\infty}^\infty dk_Q \int_0^\infty dy\, y\\
& \times  \frac{(1-t)(q_0^2+q_L^2)+(k_Q+q_Q)^2\left[(k_Q+q_Q)^2-k_Q^2\right]}{[y^2+t(1-t)(q_0^2+q_L^2)+t(k_Q+q_Q)^4+(1-t)k_Q^4]^2},\nonumber
\end{align}
where $y^2 = k_0^2+k_L^2$. The radial integral over $y$ can be performed using the radial integral identity (\ref{eqn:apxLoopC-I_radial}), 
\begin{align}
& \Pi(q)  =  \frac{g^2}{8\pi^2} \int_0^1 dt \int_{-\infty}^\infty dk_Q \\
& \times  \frac{(1-t)(q_0^2+q_L^2)+(k_Q+q_Q)^2\left[(k_Q+q_Q)^2-k_Q^2\right]}{t(1-t)(q_0^2+q_L^2)+t(k_Q+q_Q)^4+(1-t)k_Q^4}.\nonumber
\end{align}
Substituting $p=k_Q/|q_Q|$, we can write the bosonic self energy in the form 
\begin{equation}
\Pi(q) = \frac{g^2}{8\pi^2} \abs{q_Q}F\left(\frac{q_0^2 + q_L^2}{ q_Q^4}\right),
\end{equation}
where
\begin{equation}
\label{eq.Ff}
F(u) =\int_0^1 dt \int_{-\infty}^{\infty}dp \frac{(p+1)^4-p^2(p+1)^2+(1-t)u}{(p+1)^{4}t +p^4(1-t)+t(1-t)u}.
\end{equation}
This is evaluated numerically, and is used to obtain numerically exact quantum corrections.

\subsection{Soft cutoff one-loop quantum corrections}
\label{app.dLdQA2}

\noindent
Using the non-analytic RPA boson propagator 
\begin{equation}
G_\phi^{-1}(k) =  \frac{g^2}{8\pi^2} \abs{k_Q}F\bigg(\frac{k_0^2 + k_L^2}{ k_Q^4}\bigg) + m^2,
\end{equation}
without the sub-leading bare terms, we compute the loop integrals $\delta\Sigma_L$, $\delta\Sigma_Q$, and $\delta\Sigma_\Delta$, which 
arise in the expansion of the fermion self-energy correction, Eq.~(\ref{eq.fermionself_loops}), and $\delta\Xi$, which enters in the quantum correction that renormalizes
the Yukawa coupling, Eq.~(\ref{eq.otherloops}). The corresponding diagrams are shown in Figs.~\ref{fig4}(b) and (c), respectively.
The one-loop integrals we need to compute are 
\begin{align}
&\delta\Sigma_L = \frac{8\pi^2}{N_f} \Lambda\frac{d}{d\Lambda}\int_k \frac{A_k^2}{|k_Q|F_k}\left(\frac{1}{\varepsilon^2_k} - \frac{k_0^2+ k_L^2}{\varepsilon_k^4}\right),\\
&\delta\Sigma_Q = \frac{8\pi^2}{N_f} \Lambda\frac{d}{d\Lambda}\int_k  \frac{ A_k^2}{|k_Q|F_k}\times \\
&\phantom{\delta\Sigma_Q = } \times \left(\frac{(k_0^2+k_L^2)^2 - 12 (k_0^2+k_L^2)k_Q^4 + 3 k_Q^8}{\varepsilon_k^6}\right),\notag\\
&\delta\Sigma_\Delta = \frac{8\pi^2}{N_f} \Lambda\frac{d}{d\Lambda}\int_k \frac{A_k^2}{|k_Q|F_k}\left(\frac{k_0^2+k_L^2-k_Q^4}{\varepsilon^4_k}\right),\\
&\delta\Xi = -\frac{8\pi^2}{N_f} \Lambda\frac{d}{d\Lambda}\int_k \frac{A_k^3}{|k_Q|F_k}\left(\frac{1}{\varepsilon_k^2}\right),
\end{align}
where we have defined $F_k=F[(k_0^2 + k_L^2)/k_Q^4]$, $A_k = A(a_\mu\varepsilon_\mu^2(k)/\Lambda^2)$, and 
$\varepsilon^2_k=\varepsilon_\mu(k) \varepsilon_\mu(k)$, for brevity. 

\noindent
Using the transformation 
\begin{equation}
k_0 = y \cos\theta,\quad k_L = y \sin\theta ,\quad k_Q = \sqrt{y} \tilde{k}_Q,
\end{equation}
the integral over the global radial coordinate $y$ in conjunction with the logarithmic derivative $\Lambda \frac{d}{d\Lambda}$ becomes trivial due to the  
soft cutoff identity (\ref{identity}). After evaluating the angular integral over $\theta$ and substitution $u=1/\tilde{k}_Q^4$, we obtain
\begin{align}
&\delta \Sigma_L= \frac{1}{N_f}\int_{0}^{\infty}du\,\frac{1}{(1+u)^2 F(u)}=\frac{0.0797}{N_f},\\
\label{eq.SigQcomp}
&\delta \Sigma_Q= \frac{1}{N_f}\int_{0}^{\infty}du\, \frac{u^2-12u+3}{(1+u)^3 F(u)}=\frac{0.0214}{N_f},\\
&\delta \Sigma_\Delta= \frac{1}{N_f}\int_{0}^{\infty}du\, \frac{u-1}{(1+u)^2 F(u)} = \frac{0.2755}{N_f},\\
&\delta\Xi = - \frac{1}{N_f}\int_{0}^{\infty} du\, \frac{1}{(1+u)F(u)}= -\frac{0.4350}{N_f},
\end{align}
where in the last step we have numerically evaluated the integral over $u$, using the exact function $F(u)$ (\ref{eq.Ff}), which is itself a two-dimensional integral.

Alternatively it is possible to compute the corrections with the asymptotic propagator in Eq.~(\ref{eqn:apxLoopC-asymptotic_rpa}). In this case $F(u)$ is approximated by 
a closed form expression and only the one dimensional integral over $u$ needs to be performed numerically. The resulting quantum corrections are 
 $\delta\Sigma_L \approx 0.0771/N_f$, $\delta\Sigma_Q \approx 0.0250/N_f$, $\delta\Sigma_\Delta \approx 0.2759/N_f$, $\delta\Xi \approx -0.4300/N_f$.

\subsection{Soft cutoff two-loop quantum corrections}

The two-loop integrals that contribute to the mass renormalization, Eq.~(\ref{eq.otherloops}), are given by
\begin{align}
\delta \tilde\Pi &= -\frac{(8\pi^2)^2}{N_f}  \Lambda \frac{d}{d\Lambda} \int_{k,q} \frac{1}{\varepsilon_{k+q}^2 \varepsilon_k^2 \abs{q_Q}^2F_q^2} \Big[\notag\\
&\phantom{=}A_k^2 A_{k+q}^2 A_q + \frac{2 \varepsilon_{k+q}^\mu \varepsilon_k^\mu}{\varepsilon_k^2} A_k^3 A_{k+q} A_q \Big].
\end{align}

\noindent
We use the transformation 
\begin{equation}
\begin{gathered}
q_0 = y \hat{q}_0,\quad q_L = y \hat{q}_L, \quad q_Q = \sqrt{y} \tilde{q}_Q,\\
k_0 = y x \hat{k}_0,\quad k_L = y x \hat{k}_L, \quad k_Q = \sqrt{y} \tilde{k}_Q.
\end{gathered}
\end{equation}
with $\hat{k}_0^2+\hat{k}_L^2=1$ and $\hat{q}_0^2+\hat{q}_L^2=1$, e.g. $\hat{k}_0=\cos\theta$, $\hat{k}_L=\sin\theta$, $\hat{q}_0=\cos\phi$, and
$\hat{q}_L=\sin\phi$.

The global radial integral over $y$ is trivial due to the soft cutoff identity (\ref{identity}) that reflects the cutoff independence. 
We then introduce the Feynman parameter $t$ (\ref{eqn:apxLoopC-I_feynman}) to render the $x$ integral radially symmetric, 
after the shift $x\hat{k}_{0,L} \rightarrow x \hat{k}_{0,L} - t \hat{q}_{0,L}$.
Evaluating the angular integrals results in 
\begin{align}
\delta \tilde{\Pi} &=-\frac{16\pi^2 }{N_f}  \int_{k_Q,q_Q} \frac{1}{\abs{\tilde{q}_Q}^2F(\tilde{q}_Q^{-4})^2}  \int_0^1 dt\int_0^\infty dx \,x  \Big\{\notag\\
&  \frac{1}{[x^2+t(1-t) +t (\tilde{k}_Q+\tilde{q}_Q)^4+(1-t) \tilde k_Q^4]^2} \notag\\
& +\frac{ 4 (1-t)[x^2 - t(1-t) + (\tilde k_Q+\tilde q_Q)^2\tilde k_Q^2]}{[x^2+t(1-t) +t (\tilde{k}_Q+\tilde{q}_Q)^4+(1-t) \tilde k_Q^4]^3}\Big\}.
\end{align}
Using the identity (\ref{eqn:apxLoopC-I_radial}) the radial $x$ integral is evaluated resulting in
\begin{align}
\delta\tilde{\Pi} &=-\frac{2 }{N_f} \int_{-\infty}^\infty d \tilde{q}_Q \frac{1}{\abs{\tilde{q}_Q}^2 F(\tilde{q}_Q^{-4})^2} \int_0^1 dt \int_{-\infty}^\infty d\tilde{k}_Q \Big\{\notag\\
& \frac{1+2(1-t)}{[t(1-t) +t (\tilde{k}_Q+\tilde{q}_Q)^4+(1-t) \tilde k_Q^4]}\notag \\
&+ \frac{2(1-t) [(\tilde k_Q+\tilde q_Q)^2\tilde k_Q^2-t(1-t)]}{[t(1-t) +t (\tilde{k}_Q+\tilde{q}_Q)^4+(1-t) \tilde k_Q^4]^2}
\Big\}.
\end{align}
Although the integral over the Feynman parameter $t$ can be performed anaytically, we find that numerical stability of integration is enhanced if the current three-dimensional form is used.
We find that $\delta \Pi = -1.053/N_f$ with the full $F$, and $\delta \Pi \approx -1.037/N_f$  with the 
asymptotic propagator \eqref{eqn:apxLoopC-asymptotic_rpa}.

\section{Diagrams for the $\epsilon_{L,Q}$-expansions}

\subsection{Derivation of the RPA near the upper critical line}\label{app:d_L-d_Q-RPA_derivation}

Here we compute the regularized bosonic self energy (\ref{eq.regboson}) for the two cases: (i) $d_L=(3-\epsilon_L)/2,\ d_Q=1$ and (ii) $d_L=1,\ d_Q=2-\epsilon_Q$.
The first steps are carried out for general $d_L$ and $d_Q$.  Using Feynman parametrization (\ref{eqn:apxLoopC-I_feynman}) together with the 
shift $(k_0,\bm{k}_{L}) \rightarrow (k_{0},\bm{k}_{L}) - t (q_{0},\bm{q}_{L})$, the integral is 
rendered radially symmetric in the linear $(k_0, \bm{k}_{L})$ subspace, 
\begin{align}\label{eqn:APX-SD_fermion_loop}
\Pi(q) &=\frac{g^2 }{2^{d_L}\pi^{(d_L+1)/2}\Gamma(\frac{d_L+1}{2})} \int_{\bm{k}_Q}\int_0^1dt\int_0^{\infty} dy\, y^{d_L}\nonumber\\
&\frac{(1-t)(q_0^2 + \bm{q}_L^2) + (\v{k}_Q+\v{q}_Q)^2[(\v{k}_Q+\v{q}_Q)^2-\v{k}_Q^2]}{[y^2 + t(1-t) (q_0^2 + \v{q}_L^2) + t(\v{k}_Q+\v{q}_Q)^4 + (1-t)\v{k}_Q^4]^2},
\end{align}
where $k_0^2+\bm{k}_{L}^2=y^2$ and $t$ denotes the Feynman parameter. Note that we have evaluated the angular integral over the $d_L+1$ dimensional sphere and evaluated
the surface area $S_{d_L+1}$ using Eq.~(\ref{eq.volD}). Using the integral identity in Eq.~(\ref{eqn:apxLoopC-I_radial}) we can integrate over $y$, 
\begin{align}\label{eqn:APX-SD_fermion_loop_kQ}
\Pi&(q) = \frac{g^2\sec\left(\frac{d_L\pi}{2}\right)(d_L-1)}{2^{2+d_L}\pi^{(d_L-1)/2}\Gamma(\frac{d_L+1}{2})}\int_{\v{k}_Q}\int_0^1dt\bigg\{ \notag\\
&\big[(\v{k}_Q+\v{q}_Q)^2(2\v{k}_Q\cdot\v{q}_Q+\v{q}_Q^2)+(q_0^2+\v{q}_L^2)(1-t)\big]\times\notag\\
&\big[\v{k}_Q^2+(2 \v{k}_Q\cdot\v{q}_Q + \v{q}_Q^2 +(q_0^2+ \v{q}_L^2)(1-t))t\big]^{(d_L-3)/2}\bigg\}.
\end{align}

This integral cannot be computed in closed form so we look at two limits, first where $\v{q}_Q=0$, and second where $(q_0,\v{q}_L)=0$. The final asymptotic form of the propagator 
will be approximated by $G_\phi^{-1}(q) = \Pi(q_0,\v{q}_L,\v{q}_Q=0) + \Pi(q_0=0,\v{q}_L=0,\v{q}_Q)$. 
 In the first limit, $\v{q}_Q=0$, the integral does not diverge for any $d_L,d_Q>0$ and $2d_L+d_Q<6$, and results in
\begin{align}
&\Pi(q_0,\v{q}_L,\v{q}_Q=0) =\notag\\
 &- g^2 \frac{\pi^{(3-d_L-d_Q)/2}\sec\left(\frac{(2d_L+d_Q)\pi}{4}\right)}{4^{d_L+d_Q}\Gamma\left(\frac{2+d_Q}{4}\right)\Gamma\left(\frac{2d_L+d_Q}{4}\right)} (q_0^2 + \v{q}_L^2)^{\frac{2d_L+d_Q-2}{4}}.
\end{align} 

Evaluating this expression for the two starting points (i) $d_L=3/2$, $d_Q=1$ and (ii) $d_L=1$, $d_Q=2$ on the upper critical line results in the linear terms in 
Eqs. (\ref{eqn:dLdQEps-RPA_epsilonL}) and (\ref{eqn:dLdQEps-RPA_epsilonQ}).

The integral in the second limit, $(q_0,\v{q}_L)=0$, is however typically divergent on the upper critical line, but upon an evaluation in $2d_L+d_Q=4-\epsilon_{L,Q}$ we can obtain the leading $\epsilon_{L,Q}$ behavior, i.e. the $1/\epsilon_{L,Q}$ pole. This divergence can be recovered upon first making the spherical transformation $|\v{k}_Q|^4=r^2$, and then expanding the integral in the limit of large $r$ in $d_Q+2d_L=4-\epsilon_{L,Q}$. The leading term $\sim|\v{q}_Q|^{2-\epsilon_{L,Q}}/r^{1+\epsilon_{L,Q}}$ is extracted, and upon the evaluation of the integral results in 
\begin{align}
& \Pi(q_0=0,\v{q}_L=0,\v{q}_Q) =\notag \\
& \frac{(2d_L+d_Q-2)\pi^{(1-d_L-d_Q)/2}\sec\left(\frac{d_L\pi}{2}\right)}
{d_Q 2^{d_L+d_Q}(2d_L+d_Q-4)\Gamma\left(\frac{d_L-1}{2}\right)\Gamma\left(\frac{d_Q}{2}\right)}
|\v{q}_Q|^{2d_L+d_Q-2}.
\end{align}

Evaluating the pre-factor for  (i) $d_L=(3-\epsilon_L)/2,\ d_Q=1$ and (ii) $d_L=1,\ d_Q=2-\epsilon_Q$ and extracting the leading $1/\epsilon_L$ and $1/\epsilon_Q$ divergencies, 
we obtain the quadratic terms in Eqs.~(\ref{eqn:dLdQEps-RPA_epsilonL}) and (\ref{eqn:dLdQEps-RPA_epsilonQ}).

\subsection{Fermion self-energy and the Vertex correction}\label{app:dL-dQ-epsilon_corrections}

We proceed to compute the one-loop diagrams in Figs.~\ref{fig4}(b) and (c), using the soft cutoff approach. Expanding the fermion self energy diagram to leading order in external 
frequency and momenta, we obtain the quantum corrections to the linear and quadratic momentum directions as well as to the Yukawa vertex for a general $d_L$-$d_Q$ system,
\begin{align}
\delta\Sigma_L &= \frac{g^2}{N_f} \Lambda\frac{d}{d\Lambda}\int_k \left(\frac{1}{\varepsilon^2_k} - \frac{2 (k_0^2+\v{k}_L^2)}{(d_L+1)\varepsilon_k^4}\right)G_\phi(k)A_k^2,\label{eqn:apx-epsilon_sigmaL}\\
\delta\Sigma_Q &= \frac{g^2}{N_f} \Lambda\frac{d}{d\Lambda}\int_k\Bigg\{\frac{G_\phi(k)A_k^2}{\varepsilon_k^6}, \notag\\
&\bigg(\frac{4\v{k}_Q^8 - 12(k_0^2+\v{k}_L^2)\v{k}_Q^4}{d_Q}+(k_0^2+\v{k}_L^2)^2 - \v{k}_Q^8\bigg)\Bigg\},\label{eqn:apx-epsilon_sigmaQ}\\
\delta\Xi &= - \frac{g^2}{N_f}\Lambda \frac{d}{d\Lambda}\int_k \frac{G_\phi(k)A_k^3}{\varepsilon^2_k}\label{eqn:apx-epsilon_xi}.
\end{align}
Here $G_\phi(k)$ is the IR order parameter propagator defined in Eq.~(\ref{eqn:dLdQEps-RPA_epsilonL}) for the $\epsilon_L$ expansion and in 
Eq.~(\ref{eqn:dLdQEps-RPA_epsilonQ}) for the $\epsilon_Q$ expansion.

\subsubsection{$\epsilon_L$-expansion}

We compute the above integrals in $d_L=3/2,\ d_Q=1$ dimensions with the boson propagator $G_\phi(k)$ in Eq.~(\ref{eqn:dLdQEps-RPA_epsilonL}). Defining the radial coordinate $y$ in the 
$d_L+1=5/2$ dimensional $(k_0,\v{k}_L)$ subspace, $k_0^2+\v{k}_L^2=y^2$, and substituting $k_Q=\sqrt{y}x$, the $y$ integrals can be evaluated with the soft cutoff identity (\ref{identity}),
reflecting the cutoff independence. The angular integral simply gives a factor ${S}_{5/2} = (8 \pi^{5/4})/\Gamma(1/4)$. The remaining one-dimensional $x$ integrals can be computed analytically. 
Keeping the leading $\sim\sqrt{\epsilon_L}$ and first sub-leading $\sim\epsilon_L$ contributions for small $\epsilon_L$, we obtain the quantum corrections
\begin{eqnarray}
\delta\Sigma_L & = &  \frac{64\epsilon_L}{5\pi N_f}\int_0^\infty  \frac{(1+5x^4)dx}{(1+x^4)^2\left(16\sqrt{2}\ x^2 + \sqrt{\pi}\ \Gamma\left(\frac{1}{4}\right)^2\epsilon_L  \right)}\nonumber\\
& = &  \frac{2^{3/4}}{5 \pi^{1/4} \Gamma\left(\frac{5}{4}\right)}\frac{\sqrt{\epsilon_L}}{N_f},
\end{eqnarray}

\begin{eqnarray}
\delta\Sigma_Q & = & \frac{64\epsilon_L}{\pi N_f}\int_0^\infty  \frac{(1-12x^4+3x^8)dx}{(1+x^4)^3\left(16\sqrt{2}\ x^2 + \sqrt{\pi}\ \Gamma\left(\frac{1}{4}\right)^2\epsilon_L  \right)}\nonumber\\
& = & \frac{2^{3/4}}{\pi^{1/4}\Gamma\left(\frac{5}{4}\right)}\frac{\sqrt{\epsilon_L}}{N_f}-\frac{3\epsilon_L}{N_f},
\end{eqnarray}

\begin{eqnarray}
\delta\Xi & = &  -\frac{64\epsilon_L}{\pi N_f}\int_0^\infty  \frac{dx}{(1+x^4)\left(16\sqrt{2}\ x^2 + \sqrt{\pi}\ \Gamma\left(\frac{1}{4}\right)^2\epsilon_L \right)}\nonumber\\
& = & - \frac{2^{3/4}}{\pi^{1/4}\Gamma\left(\frac{5}{4}\right)}\frac{\sqrt{\epsilon_L}}{N_f} + \frac{\epsilon_L}{N_f}.
\end{eqnarray}

\subsubsection{$\epsilon_Q$-expansion}

For the expansion in the number of quadratic dimensions, we compute the integrals in Eqs.~(\ref{eqn:apx-epsilon_sigmaL})-(\ref{eqn:apx-epsilon_xi}) in $d_L=1,\ d_Q=2$,
using the IR boson propagator given in Eq.~(\ref{eqn:dLdQEps-RPA_epsilonQ}). Defining $k_0^2+k_L^2=y^2$ and $\v{k}_Q^2=y \v{x}^2$, the $y$ integral and the angular integrals 
are again trivial. Keeping the leading $\sim \epsilon_Q \log\epsilon_Q$ and first sub-leading $\sim\epsilon_Q$ contributions for small $\epsilon_Q$, the final $x$ integrals 
result in 

\begin{eqnarray}
\delta\Sigma_L & = &   \frac{16\epsilon_Q}{N_f}\int_0^\infty dx \frac{x^5}{(1+x^4)^2(8x^2+\pi^2\epsilon_Q)} \nonumber\\
& = &  \frac{\epsilon_Q}{2 N_f},\\
\delta\Sigma_Q & = &  \frac{16 \epsilon_Q}{N_f} \int_0^\infty dx \frac{x(1-6x^4+x^8)}{(1+x^4)^3(8x^2 + \pi^2 \epsilon_Q)}\nonumber\\
& = &  -\frac{\epsilon_Q}{N_f}\log\left(\frac{\pi^2\epsilon_Q}{8}\right)-\frac{2\epsilon_Q}{N_f},\\
\delta\Xi & = & - \frac{16\epsilon_Q}{N_f}\int_0^\infty dx \frac{x}{(1+x^4)( 8x^2+\pi^2\epsilon_Q)}\nonumber\\
 & = & \frac{\epsilon_Q}{N_f} \log\left(\frac{\pi^2\epsilon_Q}{8}\right).
\end{eqnarray}

\section{Comparing scaling and critical exponents}
\label{app.scaling_comp}

When comparing to critical exponents found in the literature one must be careful about the variation in definitions of the number of fermion components $N_f$, and the unit length scale $z_L=1$ or $z_Q=1$. We discuss how to do so here.

Throughout the literature, various $n$-component fermions are considered, depending on the symmetry of the initial Hamiltonian. For analytic control, the generalization to $N_n$ flavors is made. The conversion to our convention is then $N_f = n N_n$.

We have defined a unified scaling relying on the ``unit length'' $z_L \delta \ell$,
\begin{equation}
X(k) = X^{\prime}(k^{\prime})e^{-\Delta_{X}z_L\delta\ell/2},
\end{equation}
where $X = \Psi,\phi$ and $\Delta_X = [X^\dagger X] + \eta_X$ are the total scaling dimensions. Here $\eta_X$ contains all order $1/N_f$ corrections by definition. 
We did so because in the literature there are variations in the definition of the unit length scale, either using: (i) linear $z_L=1$ and (ii) quadratic $z_Q=1$  momentum directions. 
As we found in the main text, the ratio $z_L/z_Q$ renormalizes and so, for example, fixing $z_L=1$ causes $z_Q$ to renormalize with $1/N_f$ corrections.

We seek to compare to previous results in the literature, where in general the scaling is defined as 
\begin{equation}
X(k) = X^{\prime}(k^{\prime})e^{-\tilde{\Delta}_{X}\delta\ell/2},
\end{equation}
where $\tilde{\Delta}_{X} = [\widetilde{X^\dagger X}] + \tilde{\eta}_X$, 
and typically either $z_L=1$ is fixed, or $z_Q=1$ is fixed.
When the linear momentum has been defined as the unit length scale, $\tilde{\Delta}_{X} = \Delta_X$, as $z_L=1$ is fixed.
Where as, for the quadratic momentum defining the unit length scale, $\tilde{\Delta}_{X} =z_L \Delta_X$, as $z_Q=1$ but $z_L$ is not fixed.

In the case $z_Q=1$ there are subtleties in the conversion between anomalous dimensions $\eta_X$ and $\tilde{\eta}_X$.
To leading order in $N_f$ we define $z_L = z_L^{(0)} + z_L^{(1)}/N_f$, then expand $\tilde{\Delta}_{X} =z_L \Delta_X$ and equate $1/N_f$ terms resulting in the relation
\begin{equation}
\eta_X = \frac{\tilde{\eta}_X}{z_L^{(0)}} - \frac{ z_L^{(1)} [\widetilde{X^\dagger X}] }{(z_L^{(0)})^2 N_f}.
\end{equation}

There can be other variations in scaling definitions, for example Ref.~\cite{sur_roy_semi_dirac} defined
$\tilde{\Delta}_{X} =  [\widetilde{X^\dagger X}] + 2 \hat{\eta}_X$, such that $\tilde{\eta}_X = 2 \hat{\eta}_X$.
We use the comparison of the order parameter anomalous dimension as a relevant example. We also take this opportunity to correct a mistake in Table~I of Ref.~\cite{sur_roy_semi_dirac}.

In Ref.~\cite{sur_roy_semi_dirac}, $N_b$-component order parameters where coupled to $N_4$ flavors of 4-component anisotropic fermions. 
Implementing their quantum corrections (S19, S23, S26, S34) in their field-theoretic RG equations (S14) of their supplementary material finds
\begin{equation}
\hat{\eta}_\phi =\frac{\epsilon_Q}{N_4} \left[\frac{2-5 N_b}{2} + (1-N_b)\log\left(\frac{\pi^2 \epsilon_Q}{8}\right)  \right].
\end{equation} 
For the Ising case, $N_b=1$, the $\epsilon_Q \log \epsilon_Q$ correction vanishes from $\hat{\eta}_\phi$. 
This is not in contradiction to its presence in $\eta_\phi$ of Table~\ref{table:epsilon_exponents} in the main text, which was calculated with our soft cutoff methodology.
This can be clarified using the conversion 
\begin{equation}
\eta_\phi = \frac{2 \hat{\eta}_\phi}{z_L^{(0)}} + \frac{8 z_L^{(1)} }{(z_L^{(0)})^2 N_f},
\end{equation}
where $z_L$ is defined in Table~\ref{table:epsilon_exponents}, for $z_Q=1$. 
Indeed, we obtain the leading $\epsilon_Q \log \epsilon_Q$ correction.

\end{document}